\pdfoutput=1

\documentclass[12pt]{article}
\usepackage{jheppub}

\usepackage{amsmath}
\usepackage{amssymb}
\usepackage{graphicx,color,slashed,braket}
\usepackage{bbm}
\usepackage{pifont}
\usepackage{ifpdf}
\usepackage{comment}
\usepackage{array}
\usepackage{float}
\newcommand{\cN}{\mathcal{N}}
\newcommand{\cO}{\mathcal{O}}

\newcommand{\be}{\begin{equation}}
\newcommand{\ee}{\end{equation}}
\newcommand{\ba}{\begin{eqnarray}}
\newcommand{\ea}{\end{eqnarray}}
\newcommand{\nn}{\nonumber}
\newcommand{\lp}{\left(}
\newcommand{\rp}{\right)}

\newcommand{\cG}{\mathcal{G}}

\newcommand{\dd}{\mathrm{d}}

\newcommand{\GL}{\mathrm{GL}}

\newcommand{\vol}{\mathrm{dvol}}

\newcommand{\w}{\wedge}

\newcommand{\R}{\mathbb{R}}

\newcommand{\cV}{\mathcal{V}}

\newcommand{\cL}{\mathcal{L}}

\DeclareMathOperator{\Tr}{Tr}

\title{$G_2$ flux compactifications}
\author{Aravind Aikot$^a$, Zheng Miao$^a$, George Tringas$^a$, and Timm Wrase$^a$}
\affiliation{$^a$Department of Physics, Lehigh University,\\ 16 Memorial Drive East, Bethlehem, PA 18018, USA}
\emailAdd{ara425@lehigh.edu, zhm323@lehigh.edu, georgios.tringas@lehigh.edu, timm.wrase@lehigh.edu}

\abstract{We derive the three-dimensional $\mathcal{N}=1$ effective theories obtained by compactifying all five ten-dimensional string theories on generic seven-dimensional manifolds with $G_2$ structure. The resulting flux compactifications are worked out explicitly, including the full moduli dependence of the scalar potential, kinetic terms, axionic sectors, gauge fields, St\"uckelberg couplings, and the allowed geometric and form-flux data. Our results extend previous analyses by incorporating fields and fluxes that are generically present in $G_2$ reductions, and provide a unified framework for comparing type IIA, type IIB, type I and heterotic compactifications to three dimensions. In particular, the effective theories organize naturally in terms of the real superpotential formulation of three-dimensional $\mathcal{N}=1$ supergravity, making the relation between fluxes, torsion, Chern--Simons data, and moduli potentials manifest.
}
\begin{document}

\maketitle

\section{Introduction}
\label{sec:introduction}

Flux compactifications provide one of the most explicit frameworks for generating potentials for moduli fields and for realising lower-dimensional vacua of string theory with (partially) stabilised scalar sectors.
In recent years, flux compactifications to three dimensions have attracted interest as a comparatively unexplored setting for studying moduli stabilisation in AdS$_3$ and investigating the possibility of scale separation.
The aim is to understand whether their behavior parallels that of four-dimensional models or instead offers genuinely new insights into these questions.
Within this landscape, compactifications on seven-dimensional manifolds admitting $G_2$ holonomy, or more generally a $G_2$-structure, are particularly natural, since they yield minimal supersymmetry in three dimensions and give rise to low-energy effective theories with rich vacuum structures.

A first systematic derivation of three-dimensional $\cN=1$ flux vacua of massive type IIA supergravity on $G_2$ orientifolds with O2/O6-planes was presented in \cite{Farakos:2020phe}, including no-scale Minkowski vacua and classically controlled, scale-separated AdS$_3$ solutions.
The type IIB side, involving O5/O9-planes (and, for O9-planes, the inclusion of D9-branes and an $\mathrm{SO}(32)$ open-string sector as in type I), was developed in \cite{Emelin:2021gzx}.
Complementary approaches based on the bispinor formalism and other extensions \cite{VanHemelryck:2022ynr,Farakos:2023nms,Farakos:2023wps,Farakos:2025bwf} have further clarified the structure of such AdS$_3$ solutions in massive type IIA. 
However, the construction of type IIB solutions exhibiting scale separation and moduli stabilisation revealed the broader potential of three-dimensional models compared with their four-dimensional counterparts, both in constructions arising from $G_2$ compactifications and in broader approaches \cite{Arboleya:2024vnp,VanHemelryck:2025qok}.
This line of work was further developed in \cite{Arboleya:2025tax}, and complemented by the study of open-string effects in AdS$_3$ vacua \cite{Arboleya:2025lwu}.

In parallel, a program of dualities for AdS$_3$ vacua, starting from massive type IIA solutions, revealed that moduli stabilisation and scale separation can arise in controlled constructions across all ten-dimensional supergravities.
In particular, the massive type IIA constructions of \cite{Farakos:2025bwf} can be related by three T-dualities to type IIB reductions with O5/O9-planes, corresponding to the type I frame, and subsequently by S-duality to heterotic $\mathrm{SO}(32)$ compactifications.
This duality chain was explored in \cite{Miao:2025tdu}, where the type I perspective led to new parametric families of controlled AdS$_3$ vacua.
Motivated by this construction, the first parametrically controlled, scale-separated AdS$_3$ vacua on the heterotic side were constructed in \cite{Tringas:2025bwe} using $H$-flux and smeared gravitational instanton effects, potentially providing a way to study such setups from the worldsheet perspective.
Further double T-dualities lead to several families of massless type IIA solutions \cite{Tringas:2026ncg}, offering a potential complementary perspective from M-theory.
More recently, scale-separated constructions with extended supersymmetry were found in \cite{Cribiori:2026caf}, in contrast with the four-dimensional cases, where scale separation is conjectured to occur only in minimally supersymmetric or non-supersymmetric setups.

Thus, the fact that AdS$_3$ solutions with moduli stabilisation and scale separation can arise in a broader range of settings than their four-dimensional counterparts, and can therefore provide complementary ways to understand these features in string theory, further motivates the present work, in which we provide a more complete picture of these constructions.

\subsubsection*{Motivation and scope}
Despite this progress, much of the existing literature on string compactifications on $G_2$ spaces still focuses on special, highly symmetric examples, most notably toroidal orbifolds, and typically neglects resolution or blow-up modes.
In these constructions, one usually retains only a restricted set of metric deformation moduli, often corresponding to manifolds with $b^3(X)=b^4(X)=7$, where $b^k(X)$ denotes the $k$th Betti number of the internal space $X$.
As a result, the axionic and gauge sectors associated with harmonic two- and five-forms are absent; equivalently, the corresponding Betti numbers $b^5(X)=b^2(X)$, are not included in the analysis.
Moreover, while consistent truncations on $G_2$-structure spaces, for example to left-invariant forms, provide a systematic way to incorporate geometric fluxes, fully generic three-dimensional actions, including the complete set of scalar fields obtained after dualising all 3D vectors, have not yet been formulated in a single framework applying uniformly to type IIA, type IIB/type I, and heterotic supergravity.
The purpose of this paper is to fill this gap.

We study flux compactifications of ten-dimensional string theories on generic $G_2$-holonomy manifolds and on $G_2$-structure spaces with geometric fluxes, implemented through consistent truncations, while excluding non-geometric fluxes.
Our focus is on type IIA, type IIB, and type I obtained by including the relevant O9-plane involution, together with the heterotic string, where we treat the $\mathrm{SO}(32)$ and $E_8\times E_8$ heterotic theories on the same footing at the level of the universal bosonic action and its reduction.
We do not consider M-theory compactifications of $G_2$ spaces in this work.
M-theory flux compactifications on $G_2$-holonomy and $G_2$-structure manifolds have already been studied extensively, from the perspective of flux superpotentials and moduli fixing \cite{Gukov:1999ya,Beasley:2002db,Acharya:2002kv}, supersymmetry conditions and torsion classes \cite{BehrndtJeschekGstructures2003,KasteMinasianTomasiello2003,BehrndtJeschekSuperpotential2004,House:2004pm}, and effective Scherk--Schwarz/geometric-flux reductions and vacuum classifications \cite{DallAgata:2005zlf, BehrndtCveticLiu2006, McOrist:2012yc}.
For a recent weak $G_2$ and scale-separation perspective see~\cite{VanHemelryck:2024bas}.

Our main objective is to derive the fully generic three-dimensional $\cN=1$ effective actions for the above string theories compactified on $G_2$ spaces, keeping the dependence on the $G_2$ geometry as explicit as possible. While we will highlight representative consequences for moduli stabilisation and controlled AdS$_3$ vacua, we defer a more detailed discussion of phenomenological and Swampland-related implications to the conclusions.

\medskip
\subsubsection*{Organisation of the paper}
In Section~\ref{sec:G2prelim}, we collect the relevant geometric background on $G_2$ spaces and establish our conventions.
In Section~\ref{sec:10dto3d_general}, we set up the general ten-to-three-dimensional reduction and the corresponding 3D $\mathcal{N}=1$ supergravity data.
We then derive the effective actions and flux superpotentials for type IIA and type IIB/type I reductions in Sections~\ref{sec:IIA} and~\ref{sec:IIB}, respectively, including their tadpole constraints and orientifold projections.
In Section~\ref{sec:heterotic}, we discuss heterotic string compactifications, including the role of $\alpha'$-corrected Bianchi identities and gravitational and gauge instanton effects.
Finally, in Section~\ref{sec:conclusions}, we conclude with a summary and outlook.
The appendices collect our conventions and further technical details.

\section{\texorpdfstring{Geometric preliminaries on $G_2$ manifolds}{}}\label{sec:G2prelim}

In this section, we review $G_2$ spaces and present the definitions and identities needed for our general constructions in flux compactifications.

\subsection{\texorpdfstring{$G_2$-structures and form decompositions}{}}\label{subsec:G2structures}

Let $X$ be an oriented smooth 7-manifold. Then a \emph{$G_2$-structure} on $X$ is defined by a smooth positive 3-form $\Phi$, meaning that at each point of $X$, the form $\Phi$ lies in the open $\GL(7,\mathbb{R})$-orbit of the standard associative 3-form on $\mathbb{R}^7$. Such a form $\Phi$ determines a Riemannian metric $g=g(\Phi)$ and the corresponding Hodge star operator $\star=\star_{g(\Phi)}$. In particular, it also defines the associated \emph{coassociative} 4-form
\begin{equation}\label{eq:defpsi}
  \Psi \equiv \star \Phi \,.
\end{equation}
Here $\Phi\in\Omega^3(X)$ and $\Psi\in\Omega^4(X)$, where $\Omega^p(X)$ denotes the space of smooth differential $p$-forms on $X$.

Given the Levi--Civita connection $\nabla^{(g)}$ associated with the metric determined by the $G_2$-structure, the failure of $\Phi$ to be covariantly constant, $\nabla^{(g)}\Phi \neq 0$, is encoded in the intrinsic torsion. Equivalently, the latter can be packaged into four torsion classes, see \cite{FernandezGray, 2003math......5124B},
\begin{align}
  \dd_7 \Phi &= \tau_0\,\Psi + 3\,\tau_1\wedge \Phi + \star \tau_3 \,,\label{eq:dPhiTorsion}\\
  \dd_7 \Psi &= 4\,\tau_1\wedge \Psi + \tau_2\wedge \Phi \,,\label{eq:dPsiTorsion}
\end{align}
where $\tau_0\in\Omega^0_1(X)$, $\tau_1\in\Omega^1_7(X)$, $\tau_2\in\Omega^2_{14}(X)$, and $\tau_3\in\Omega^3_{27}(X)$. Here $\Omega^p_r(X)$ denotes the subspace of $p$-forms transforming in the $r$-dimensional irreducible representation of $G_2$. Different types of $G_2$-structures are obtained by setting appropriate subsets of these torsion classes to zero. The Ricci scalar can be written in terms of the above torsion classes as \cite{2003math......5124B}
\begin{equation}
R_7 = 12 \dd_7^\dagger \tau_1 +\frac{21}{8} \tau_0^2 + 30 |\tau_1|^2 - \frac12 |\tau_2|^2 -\frac12 |\tau_3|^2\,.\label{eq:R7viaTorsion}
\end{equation}

The $G_2$-structure is \emph{torsion-free} if $\nabla^{(g)}\Phi=0$ and thus the 3-form $\Phi$ is covariantly constant with respect to the Levi--Civita connection, which is equivalent to
\begin{equation}\label{eq:torsionfree} 
  \dd_7 \Phi = 0\,, \qquad \dd_7 \Psi = 0\,.
\end{equation}
In this paper, we are assuming the torsion-free $G_2$-structure to have full $G_2$ holonomy, i.e. $\mathrm{Hol}(g)=G_2$, rather than only assuming that the holonomy is contained in $G_2$, $\mathrm{Hol}(g)\subseteq G_2$, which means we exclude other torsion-free cases where the holonomy could be contained in a proper subgroup of $G_2$ \cite{JoyceBookSpecialHolonomy}.

At each point of a $G_2$-structure manifold, the spaces of $p$-forms admit a decomposition into irreducible representations $\Lambda^p$ of $G_2$.
In particular\footnote{Pointwise $\Lambda^*_7$ is nontrivial, but on a compact torsion-free manifold with full $G_2$ holonomy there are no harmonic representatives in this component, so $H^*_7(X)=0$.},
\begin{align}
  \Lambda^2 &= \Lambda^2_{7} \oplus \Lambda^2_{14}\,,
  \label{eq:Lambda2decomp}\\
  \Lambda^3 &= \Lambda^3_{1} \oplus \Lambda^3_{7} \oplus \Lambda^3_{27}\,,
  \label{eq:Lambda3decomp}
\end{align}
and similarly for $\Lambda^p\simeq \star\Lambda^{7-p}$. A convenient characterization is given in terms of wedge products with $\Phi$ and $\Psi$, which distinguish the various irreducible components through algebraic identities involving differential forms of different degrees, see \cite{FernandezGray, JoyceBookSpecialHolonomy, KarigiannisDeformations}. We present some identities that are useful for one of the main features of this paper, namely the inclusion of two-forms and axions; for further details, we refer the reader to the literature. For example, a 2-form $\alpha$ decomposes into the irreducible $G_2$ representations $\Lambda^2_{7}$ and $\Lambda^2_{14}$, and these components are characterized by the relations
\begin{align}
  \Lambda^2_{7}
  &= \left\{ \alpha \in \Lambda^2 \;\middle|\; \star(\Phi \wedge \alpha) = 2\,\alpha
     \;\Longleftrightarrow\; \star\alpha = \tfrac12\,\alpha \wedge \Phi \right\},\label{eq:Lambda2characterization7} \\
  \Lambda^2_{14}
  &= \left\{ \alpha \in \Lambda^2 \;\middle|\; \star(\Phi \wedge \alpha) = -\,\alpha
     \;\Longleftrightarrow\; \star\alpha = -\,\alpha \wedge \Phi \right\}.\label{eq:Lambda2characterization}
\end{align}

\subsection{\texorpdfstring{Harmonic 3-forms, $G_2$ deformations and metrics}{}}
\label{subsec:G2moduli_metric3}

Assume first that $X$ is compact and has full $G_2$ holonomy, and let $\Phi_I$ be a basis of harmonic 3-forms representing $H^3(X,\mathbb{R})$, with
\begin{equation}\label{eq:iRange}
  I=1,\dots,b^3(X)\,.
\end{equation}
A standard parametrization of the $G_2$-structure is
\begin{equation}\label{eq:phiExpansion}
  \Phi=s^I\Phi_I\,,
\end{equation}
where the real scalars $s^I$ are local coordinates on the moduli space of torsion-free $G_2$-structures with full $G_2$ holonomy.
Globally, the relation between $s^I$ and periods of $\Phi$ depends on the chosen basis and on the fact that $\star$ itself depends on $\Phi$.

The metric on the space of deformations is given by
\begin{equation}\label{eq:GijDef}
  G_{IJ}=\frac{1}{\mathcal{V}}\int_X \Phi_I \wedge \star \Phi_J \,,
\end{equation}
where $\mathcal{V}\equiv \mathcal{V}(s)$ is the volume functional,
\begin{equation}\label{eq:volfunctional}
  \mathcal{V}=\frac{1}{7}\int_X \Phi\wedge \star\Phi = \int_X \mathrm{dvol}_7 \,.
\end{equation}
Here $\mathrm{dvol}_7$ denotes the Riemannian volume form on $X$ induced by $\Phi$, more explicitly, $\mathrm{dvol}_7=\sqrt{g}\, d^7y$, where $y$ are local coordinates on $X$.
The normalization by $\mathcal{V}$ is conventional; for many later formulae only the un-normalized pairing $\int \Phi_I\wedge\star\Phi_J$ matters. The metric $G_{IJ}$ is positive definite and related to the scalar field metric for the $s^I$ as we show below.

\subsection{\texorpdfstring{Harmonic 2-forms, topological couplings, and a second metric}{}}\label{subsec:G2metric2forms}

If $b^2(X)\neq 0$, expanding higher-form potentials along harmonic 2-forms gives rise to additional massless fields in the lower-dimensional compactification, typically axions and abelian gauge fields.
In contrast to the harmonic 3-forms, these modes do not parametrize geometric deformations of the torsion-free $G_2$-structure.

Let $\Upsilon_A$ with $A=1,\dots,b^2(X)$ be a basis of harmonic 2-forms representing $H^2(X,\mathbb{R})$.
On a manifold with full $G_2$ holonomy, harmonic 2-forms lie in the $\mathbf{14}$ representation, i.e.\ $\Upsilon_A\in\Lambda^2_{14}$.
In that case the identity \eqref{eq:Lambda2characterization} applies to each $\Upsilon_A$
\begin{equation}\label{eq:starUpsilon}
  \star \Upsilon_A = -\,\Upsilon_A\wedge\Phi\,.
\end{equation}
The kinetic metric for the corresponding modes will be related to the following metric
\begin{equation}
  G_{AB}=\frac{1}{\mathcal{V}}\int_X \Upsilon_A \wedge \star \Upsilon_B \,.
  \label{eq:GabDef}
\end{equation}
We now introduce the mixed intersection couplings
\begin{equation}
  \kappa_{AB I}=\int_X \Upsilon_A \wedge \Upsilon_B \wedge \Phi_I \,.
  \label{eq:kappaabiDef}
\end{equation}
These are topological in the sense that they depend only on the cohomology classes of $\Upsilon_A,\Upsilon_B,\Phi_I$.
Using \eqref{eq:phiExpansion} and the identity \eqref{eq:starUpsilon}, we can relate $\kappa_{ABI}$ to the metric-dependent kinetic matrix $G_{AB}$ in equation \eqref{eq:GabDef}
\begin{equation}\label{eq:GabFromKappa}
  G_{AB}=\frac{1}{\mathcal{V}}\int_X \Upsilon_A \wedge \star \Upsilon_B
  = -\frac{1}{\mathcal{V}} \int_X \Upsilon_A \wedge \Upsilon_B \wedge \Phi=-\frac{s^I}{\mathcal{V}}\kappa_{ABI}\,.
\end{equation}
This relation is particularly useful in compactifications that include fields arising from two-forms, since it makes the $s$-dependence of $G_{AB}$ explicit in terms of the moduli $s^I$, the volume $\mathcal{V}(s)$, and the topological data $\kappa_{ABI}$.
Equation \eqref{eq:GabFromKappa} relies on the fact that, for compact torsion-free $G_2$-manifolds with full $G_2$ holonomy, the relevant harmonic 2-forms lie in $\Lambda^2_{14}$, where the identity \eqref{eq:starUpsilon} applies.
If one considers a more general $G_2$-structure space, or a truncation to a subspace of left-invariant forms that is not harmonic, then \eqref{eq:starUpsilon} and hence \eqref{eq:GabFromKappa} can receive corrections, and one must specify the truncation scheme.
In this paper we assume that the basis of 2-forms satisfies equation \eqref{eq:starUpsilon}.

\subsection{\texorpdfstring{$G_2$-structure spaces and intrinsic torsion}{}}\label{subsec:G2torsion}

While in the $G_2$-holonomy case we expand in harmonic forms, for $G_2$-structure reductions we instead assume a consistent truncation to finite-dimensional subspaces of internal forms closed under $\dd_7$, typically spanned by left-invariant forms.
We denote bases of the truncated spaces of internal 2-forms and 3-forms by
\begin{equation}
  \{\Upsilon_A\}\subset \Omega^2(X)\,,
  \qquad
  \{\Phi_I\}\subset \Omega^3(X)\,.
\end{equation}
Note that we use the same indices $A$ and $I$, but their ranges are no longer determined by the Betti numbers $b^2$ and $b^3$.
Indeed, the forms in the truncation basis are generically non-closed, with $\dd_7 \Upsilon_A \neq 0$ and $\dd_7 \Phi_I \neq 0$, but their exterior derivatives nevertheless close on the truncated basis.
Consequently, these forms do not define cohomology classes, and their number is not determined by the cohomology of $X$.
To simplify the discussion below we refer to the number of fields or fluxes as $b^2$ or $b^3$, keeping in mind that for a $G_2$-structure manifold this should be rather the number of left-invariant 2- and 3-forms, respectively.
It will be useful to introduce dual bases of 5-forms and 4-forms,
\begin{equation}
  \{T_A\}\subset \Omega^5(X)\,,
  \qquad
  \{\Psi_I\}\subset \Omega^4(X)\,,
\end{equation}
chosen such that they realise the \emph{Poincar\'e pairings} on the truncated subspaces
\begin{equation}
  \int_X \Upsilon_A \wedge T_B = \delta_{AB}\,,
  \qquad
  \int_X \Phi_I \wedge \Psi_J = \delta_{IJ}\,.
  \label{eq:PoincareDualBases}
\end{equation}
In the $G_2$ holonomy case, when the basis forms are harmonic, these reduce to the usual Poincar\'e pairings on cohomology, $H^2(X)\times H^5(X)\to\mathbb{R}$ and $H^3(X)\times H^4(X)\to\mathbb{R}$, while in a left-invariant truncation \eqref{eq:PoincareDualBases} is understood simply as a convenient choice of dual bases within the truncated subspaces of forms.

As discussed in subsection~\ref{subsec:G2structures}, the intrinsic torsion is encoded by the failure of the truncated basis forms to be closed see \eqref{eq:dPhiTorsion}-\eqref{eq:dPsiTorsion}.
We parameterize this by constant matrices
$N_A{}^{I}$, $M_I{}^{J}$ and $Q_J{}^{A}$ defined through
\begin{align}
  \dd_7 \Upsilon_A &= N_A{}^{I}\Phi_I\,,
  \label{eq:dUpsilonMatrix}\\
  \dd_7 \Phi_I &= M_I{}^{J}\Psi_J\,,
  \label{eq:dPhiMatrix}\\
  \dd_7 \Psi_J &= Q_J{}^{A}T_A\,.
  \label{eq:dPsiMatrix}
\end{align}
These matrices are the analogues, in a form-basis language, of the structure constants on twisted tori \cite{DallAgata:2005zlf,Emelin:2021gzx,Miao:2025tdu}, and they control the geometric-flux contributions to the superpotential, the dimensionally reduced potential, and the Bianchi identities.
In the following sections, the matrices $N$, $M$, and $Q$ enter the dimensional reduction through the non-closure of the internal basis and constitute the precise data encoding what is commonly called \emph{metric flux} in the effective theory.
For example, they appear in the RR and NSNS Bianchi identities through terms such as $\dd_7(b^A\Upsilon_A)$, and they also enter the flux superpotential through the contribution associated with the non-vanishing curvature of the internal space.

The nilpotency condition $\dd_7^2=0$ on the basis forms implies algebraic constraints on the matrices and from $\dd_7^2\Upsilon_A=0$ and $\dd_7^2\Phi_I=0$ one finds
\begin{equation}\label{eq:d2constraints}
  N_A{}^{I} M_I{}^{J} = 0\,,
  \qquad
  M_I{}^{J} Q_J{}^{A} = 0\,.
\end{equation}
Moreover, integration by parts together with \eqref{eq:PoincareDualBases} relates $Q$ and $N$
\begin{equation}\label{eq:QminusNT}
  0=\int_X \dd_7(\Upsilon_A\wedge \Psi_J)
   \;\;\rightarrow\;\;
  Q_J{}^{A} = -\,N_A{}^{J}\,,
\end{equation}
i.e.\ $Q=-N^{\mathrm T}$.
Similarly,
\begin{equation}\label{eq:Msymmetric}
  0=\int_X \dd_7(\Phi_I\wedge \Phi_J)
  \;\;\rightarrow\;\;
  M_{I}{}^J=M_{J}{}^I\,,
\end{equation}
so $M_{I}{}^J$ is symmetric.

Given the associative form written in terms of the moduli in \eqref{eq:phiExpansion}, and taking the exterior derivative on it produces
\begin{equation}\label{eq:dPhiInternalTrunc}
  \dd_7\Phi = s^I\dd_7\Phi_I = s^I M_I{}^{J}\Psi_J\,.
\end{equation}

Comparing \eqref{eq:dPhiInternalTrunc} and \eqref{eq:dPsiMatrix} with the general torsion-class decomposition \eqref{eq:dPhiTorsion}--\eqref{eq:dPsiTorsion} shows that the matrices $M$ and $N$ (or $Q$) encode the components of the intrinsic torsion that survive in the truncated basis.
For example, on co-calibrated $G_2$-structures without 2- and 5-forms, in constructions such as in \cite{DallAgata:2005zlf, Danielsson:2014ria, Emelin:2021gzx, Miao:2025tdu}, one has $Q=N=0$ since $\dd_7\Psi=\dd_7\Upsilon=0$, while $M$ can still be non-zero and encodes the remaining torsion components in $\dd_7\Phi$.

%%%%%%%%%%%%%%%%%%%%%%%%%%%%%%%%%%%%%%%%%%%%%%
%%%%%%%%%%%%%%%%%%%%%%%%%%%%%%%%%%%%%%%%%%%%%%
%%%%%%%%%%%%%%%%%%%%%%%%%%%%%%%%%%%%%%%%%%%%%%

\section{\texorpdfstring{General reduction from 10D to 3D $\mathcal{N}=1$ effective theory}{}}\label{sec:10dto3d_general}

In this section, we set up a common framework for flux compactifications of type~IIA, type~IIB, type~I, and the heterotic supergravities to three dimensions on $G_2$ spaces.
Our aim is to keep the discussion as theory-independent as possible and to isolate the structural ingredients that will later be specialized to the different string theories.
These include the conventions for the ten-dimensional action, the rescaling to three-dimensional Einstein frame, and the universal form of the three-dimensional $\mathcal{N}=1$ action in terms of a scalar field metric and a real superpotential.
Explicit expressions for these quantities in the various compactifications will be presented in the following sections.

\subsection{NSNS sector and gauge form fields}
\label{subsec:10d_universal_action}

We work in ten-dimensional Einstein frame with mostly-plus signature.
The universal NSNS sector contains the metric $g_{MN}$, the dilaton $\phi$, and the Kalb--Ramond field $B_2$ with field strength $H_3 = \dd B_2$, up to Chern--Simons modifications in the heterotic/type~I theories. To treat the type II RR fields and the heterotic/type I gauge fields in a unified way, we introduce a set of $p_\Lambda$-form field strengths $\mathcal{F}^{\Lambda}_{p_\Lambda}$. Depending on the theory under consideration, these represent either the RR field strengths in type II or the gauge-field strengths in the heterotic and type I theories. The bosonic action is taken to be
\begin{align}\label{eq:S10_universal}
  S_{10}
  &= \frac{1}{2\kappa_{10}^2}\int_{M_{10}}
     d^{10}X\sqrt{-G}
     \left(
       R_{10}
       -\frac{1}{2}(\partial\phi)^2
       -\frac{1}{2}e^{-\phi}|H_3|^2
       -\sum_\Lambda\frac{1}{4}e^{a_\Lambda\phi}\bigl|\mathcal{F}^{\Lambda}_{p_\Lambda}\bigr|^2
     \right)
     + S_{\mathrm{loc}}\,,
\end{align}
where $2\kappa_{10}^2=(2\pi)^7\alpha'^4$ and $\alpha'=l_s^2$.
The action $S_{\mathrm{loc}}$ collectively denotes possible localized sources. The Dp-branes and Op-planes that can appear in type II and type I string theory settings are discussed in appendix \ref{app:AppendixII}.

Our convention for the norm of a $p$-form is
\begin{equation}
  |F_p|^2 = \frac{1}{p!}\,F_{M_1\ldots M_p}F^{M_1\ldots M_p}\,,
  \qquad
  \int F_p\wedge \star F_p = \int \sqrt{-g}\,|F_p|^2\,.
  \label{eq:formnorm}
\end{equation}
and the fluxes are quantized in the following way
\begin{equation}
    \int_{\Sigma_p} F_p=\big(2\pi\sqrt{\alpha^{\prime}}\big)^{p-1}N_p\,,\qquad N_p\in \mathbb{Z}\,,
\end{equation}
where $N_p$ is the relevant flux quantum and $\Sigma_p \in H_p(X,\mathbb{Z})$. In this paper we work in units where $2\pi\sqrt{\alpha^{\prime}}=1$, except in the heterotic string section where we find it useful to keep explicit factors of $\alpha'$.

In later sections we will specify $S_{\mathrm{loc}}$ and the values of $a_\Lambda$ appropriate to each string theory.
\begin{itemize}
  \item In type~II/I, the RR kinetic terms in Einstein frame take the form \eqref{eq:S10_universal} with
  \begin{equation}
    a_{\Lambda}=\frac{5-p}{2}\,,
    \label{eq:aRR}
  \end{equation}
  where $p$ is the RR field-strength degree.
  More specifically, $p$ is even in IIA and odd in IIB. We will work in the democratic formulation where one additionally imposes duality constraints at the level of the equations of motion. We discuss details of compactifications in those theories in Sections~\ref{sec:IIA}-\ref{sec:IIB}.
  \item In the heterotic and type~I theories (at leading order in $\alpha'$), the gauge field strength is a 2-form $F_2$ and its Einstein-frame kinetic term corresponds to \eqref{eq:S10_universal} with respectively
  \begin{equation}
    a^{(\mathrm{YM})}_{\rm het} = -\frac12\,, \qquad a^{(\mathrm{YM})}_{\rm type\, I} = \frac12\,.
    \label{eq:aYM}
  \end{equation}
  Moreover, $H_3$ receives the Green--Schwarz modification and we defer its detailed treatment, and the higher-derivative $\alpha'$ corrections, to the heterotic Section~\ref{sec:heterotic}.
\end{itemize}
For the general structure of the ten-dimensional actions and these dilaton couplings we follow standard references such as \cite{BeckerBeckerSchwarzBook}.

\subsection{Metric ansatz, truncation, and the 3D Einstein frame}\label{subsec:3d_Einstein_frame}

We compactify the ten-dimensional supergravity action in \eqref{eq:S10_universal} on a seven-dimensional internal space $X$ with $G_2$ structure space and assume a product topology
\begin{equation}
  M_{10} = M_{2,1}\times X\,.
  \label{eq:product}
\end{equation}
To derive the three-dimensional effective theory, we consider the following metric ansatz
\begin{equation}
  \dd s^2_{10}
  =
  \mathcal{V}^{-2}g_{\mu\nu}\dd x^\mu \dd x^\nu
  +
  g_{mn}\dd y^m \dd y^n\,,
  \label{eq:metricAnsatz_Einstein3d}
\end{equation}
where $g_{\mu\nu}\equiv g_{\mu\nu}(x)$, and $g_{mn}\equiv g_{mn}(y,s(x))$ denotes the dimensionless metric on the internal $G_2$ manifold with corresponding volume \eqref{eq:volfunctional}.

The prefactor of the external metric in \eqref{eq:metricAnsatz_Einstein3d} is chosen so that we are automatically in the 3d Einstein frame and do not have to perform a Weyl rescaling to remove the factor of the internal volume that arises when integrating the 10d action over the internal space. Let us briefly neglect the kinetic terms for the scalar fields to show this explicitly by reducing equation \eqref{eq:S10_universal} 
\begin{align}
    S_{10} &\supset 2\pi \int_{M_{10}}
     d^{10}X\sqrt{-G}\left(R_{10} -\frac{1}{2}e^{-\phi}|H_3|^2-\sum_\Lambda\frac{1}{4}e^{a_\Lambda\phi}|\mathcal{F}^{\Lambda}_{p_\Lambda}|^2\right) + S_{\mathrm{loc}}\nonumber \\
     &\supset 4\pi \int_{M_{2,1}}d^{3}x\sqrt{-g_3}\ \mathcal{V}^{-3}\left(\frac{1}{2}\int_X d^{7}y\sqrt{g_7}\left(\mathcal{V}^2R_3+R_7\right)-\mathcal{V}^3 \, V_{\text{flux}}-\mathcal{V}^3 \, V_{\text{loc}}\right)  \nonumber \\
     &=4\pi \int_{M_{2,1}}
     d^{3}x\sqrt{-g_3}\left(\frac{1}{2}R_{3}
     -V_{\text{curv}}
     -V_{\text{flux}}
     -V_{\text{loc}}\right)  \,.
\end{align}
The three-dimensional scalar potential is generically given by
\begin{align}
    V&=V_{\text{curv}} + V_{\text{flux}} + V_{\text{loc}} \,,
\end{align}
with
\begin{align}
    V_{\text{curv}}&=-\frac{1}{2\mathcal{V}^3}\int_Xd^7y \sqrt{g_7}R_7\,,\\
    V_{\text{flux}}&=\frac{1}{2\mathcal{V}^3}\int_X d^7y\sqrt{g_7}\lp \frac{1}{2}e^{-\phi}|H_3|^2 +\sum_\Lambda\frac{1}{4}e^{a_\Lambda\phi}|\mathcal{F}^{\Lambda}_{p_\Lambda}|^2\rp\,,
\end{align}
and $V_{\mathrm{loc}}$ arises from reducing $S_{\mathrm{loc}}$. 

The dilaton kinetic term trivially descends from the ten-dimensional Einstein-frame action in equation \eqref{eq:S10_universal} and we find explicitly 
\begin{equation}
  (4\pi e)^{-1}\cL_{\phi}
  =
  -\frac{1}{4}(\partial\phi)^2\,.
  \label{eq:DilatonKinetic}
\end{equation}
where $e\equiv \sqrt{-g_3}$.

The kinetic terms for the $G_2$ moduli receive two contributions: one from the dependence of the internal metric on the external coordinates, that is, from the reduction of the internal Einstein--Hilbert term, and another from the Weyl factor $\cV(x)^{-2}$ multiplying the external metric in the Ansatz \eqref{eq:metricAnsatz_Einstein3d}. We present the detailed derivation of the kinetic term in Appendix \ref{app:sIkinetic} and state here the answer
\begin{equation}\label{eq:G2moduliMetricPhysicalLagrangian}
  (4\pi e)^{-1}\,\mathcal L_{s}
  =
  -\mathcal G_{IJ}\partial_\mu s^I\partial^\mu s^J\,,
\end{equation}
with
\begin{equation}\label{eq:G2moduliMetricPhysicalExplicit}
  \mathcal G_{IJ} = \frac14G_{IJ} + 
  \frac{1}{36\cV^2}
  \left(\int_X \Phi_I\wedge\Psi\right)
  \left(\int_X \Phi_J\wedge\Psi\right) \,,
\end{equation}
where $G_{IJ}=\frac{1}{\mathcal{V}}\int_X \Phi_I \wedge \star \Phi_J$ from above in equation \eqref{eq:GijDef}.

The remaining ten-dimensional fields are expanded on a finite basis of internal forms.
For instance, the Kalb--Ramond field may be expanded as
\begin{equation}\label{eq:B2exp_generic}
  B_2(x,y)=b^A(x)\Upsilon_A(y) + \dots\,,
\end{equation}
where $\{\Upsilon_A\}$ is the basis of internal 2-forms from above, the ellipsis denotes additional components, including possible purely external forms, and the $b^A(x)$ are three-dimensional axions. Similarly, for a generic ten-dimensional gauge potential one may write schematically
\begin{equation}\label{eq:ApExp_generic}
  \mathcal{A}^{\Lambda}_{p_\Lambda-1}(x,y)
  =
  \sum_{n} a^{\Lambda,n}(x)\omega^{\Lambda}_{p_\Lambda-1,n}(y)+ \dots\,,
\end{equation}
where the coefficients $a^{\Lambda,n}(x)$ become three-dimensional scalars, axions, or vectors, depending on how many external legs are carried by the corresponding ten-dimensional field component.

The kinetic terms of the lower-dimensional fields follow by inserting the expansions into the ten-dimensional action and integrating over the internal space.
For the $B$-field, one has
\begin{equation}\label{eq:H3split_generic}
  H_3
  =
  H_3^{\mathrm{bg}}
  +
  \dd_3 b^A \wedge \Upsilon_A
  +
  b^A\dd_7 \Upsilon_A\,,
\end{equation}
where $\dd_3$ and $\dd_7$ denote the external and internal exterior derivatives, respectively, and $H_3^{\mathrm{bg}}$ is the background internal flux.
The part with one external leg, $\dd_3 b^A \wedge \Upsilon_A$, gives rise to the scalar kinetic terms
\begin{equation}\label{eq:kineticb_generic}
  S_{\mathrm{kin}}[b]
  =-4\pi\int_{M_{2,1}}\!\!d^3x\sqrt{-g_3}\,\mathcal{G}_{AB}\partial_\mu b^A\partial^\mu b^B\,,
\end{equation}
with a field-space metric, $\mathcal{G}_{AB}\equiv\mathcal{G}_{AB}(\phi,s)$, obtained from an internal overlap integral,
\begin{equation}\label{eq:kineticmetricb_generic}
  \mathcal{G}_{AB}
  =\frac{e^{-\phi}}{4 \mathcal{V}} 
  \int_X\Upsilon_A \wedge \star \Upsilon_B = \frac{1}{4} e^{-\phi} G_{AB} \,,
\end{equation}
where we have assumed that the dilaton does not depend on the internal coordinates.
Exactly the same logic applies to the coefficients $a^{\Lambda,n}(x)$ in equation \eqref{eq:ApExp_generic}.
Considering that the corresponding field strength contains a piece of the form
\begin{equation}
  \mathcal{F}^{\Lambda}_{p_\Lambda}
  \supset
  \dd_3 a^{\Lambda,n}\wedge \omega^{\Lambda}_{p_\Lambda-1,n}\,,
\end{equation}
then the flux term in \eqref{eq:S10_universal} produces a three-dimensional kinetic term
\begin{equation}\label{eq:kinetica_generic}
  S_{\mathrm{kin}}[a^\Lambda]
  =
  -4\pi \int_{M_{2,1}}\!\!d^3x\sqrt{-g_3}\,
  \mathcal{G}^{(\Lambda)}_{nm}\,\partial_\mu a^{\Lambda,n}\,\partial^\mu a^{\Lambda,m}\,,
\end{equation}
with
\begin{equation}\label{eq:kineticmetrica_generic}
  \mathcal{G}^{(\Lambda)}_{nm}
  =
  \frac{1}{8 \mathcal{V}} \int_X e^{a_\Lambda\phi}\,
  \omega^{\Lambda}_{p_\Lambda-1,n}\wedge \star\omega^{\Lambda}_{p_\Lambda-1,m}\,,
\end{equation}
and $\mathcal{G}^{(\Lambda)}_{nm}\equiv\mathcal{G}^{(\Lambda)}_{nm}(\phi,s)$. Note that in the democratic formalism for type II/I below the RR-field strengths $\widetilde{F}_p$ and their duals $\star_{10} \widetilde{F}_{10-p}$ are identified, leading to an effective factor of 2 in the above kinetic term.

The same ten-dimensional terms also generate the lower-dimensional scalar potential.
Indeed, the purely internal pieces of the field strengths,
$H_{3}^\mathrm{int} = b^A\,\dd_7\Upsilon_A + H_3^{\mathrm{bg}}$ in \eqref{eq:H3split_generic},
contribute through internal overlap integrals of the form
\begin{equation}
  V_H
  \;=\;
  \frac{1}{4 \mathcal{V}^3}\,
  \int_X e^{-\phi}\,
  H_{3}^\mathrm{int}\wedge \star H_{3}^\mathrm{int}\,,
  \label{eq:VH_generic}
\end{equation}
and similarly for the internal parts of the $\mathcal{F}^{\Lambda}_{p_\Lambda}$.
In particular, background flux, non-closure of the basis forms (metric flux), and internal axion-dependent flux combinations
all contribute to the scalar potential.
The internal curvature term obtained from $R_{10}$ likewise produces both kinetic terms for the geometric moduli
and a contribution to the scalar potential whenever the internal space is not Ricci-flat, as is generically the case
for $G_2$-structure compactifications with intrinsic torsion.
Thus the ten-dimensional curvature and flux terms are responsible for both the scalar kinetic terms and the scalar potential of the three-dimensional effective theory.

\subsection{\texorpdfstring{Universal form of the 3D $\mathcal{N}=1$ effective action}{}}\label{subsec:3dN1_general}

With the orientifold/projection conditions typically imposed in type~II compactifications on $G_2$ spaces, and for the minimal supersymmetric heterotic $G_2$ systems, the resulting low-energy theory in three dimensions has minimal supersymmetry, i.e.\ $\mathcal{N}=1$ with two real supercharges.
A major simplification in three dimensions is that abelian vectors can be dualized to scalars, so the light field content can be organized into real scalar multiplets.

The general two-derivative $\mathcal{N}=1$ supergravity coupled to real scalars takes the form \cite{Emelin:2021gzx}
\begin{equation}\label{eq:L3_generic}
  (4\pi e)^{-1}\mathcal{L}_{3}
  =
  \frac12 R_3
  -
  \mathcal{G}_{MN}\partial_\mu\varphi^M\partial^\mu\varphi^N
  -
  V
  +\dots\,,
\end{equation}
where $\varphi^M$ collectively denotes all real scalar fields:
geometric moduli, dilaton/volume combinations, axions descending from $B_2$ and the RR/gauge potentials, and dual scalars obtained from three-dimensional abelian vectors.
The ellipsis indicates possible non-abelian vector fields, topological couplings and fermionic terms.

The details of the block diagonal scalar field metric depend on the particular string theory compactification but there will always be a term $1/4$ for the dilaton from equation \eqref{eq:DilatonKinetic}, a term for the $s^I$ moduli from equation \eqref{eq:G2moduliMetricPhysicalExplicit} and kinetic terms for $B_2$ axions from equation \eqref{eq:kineticmetricb_generic} as well as potentially other axion kinetic terms from equation \eqref{eq:kineticmetrica_generic}. So, we schematically have\footnote{For type IIB below, there is kinetic mixing between $C_4$ and $B_2$ axions, leading to off-diagonal kinetic terms $\mathcal{G}_{Am}$.}
\begin{equation}
    \mathcal{G}_{MN} = \begin{pmatrix}
\frac14 & 0 & 0 & 0\\
0 & \mathcal{G}_{IJ} & 0 & 0\\
0 & 0 & \mathcal{G}_{AB} & 0\\
0 & 0 & 0 & \mathcal{G}^{(\Lambda)}_{nm}\\
\end{pmatrix}\,,
\end{equation}
where $\mathcal{G}_{MN}\equiv \mathcal{G}_{MN}(\varphi)$.

Supersymmetry fixes the three-dimensional scalar potential $V\equiv V(\varphi)$ in terms of a \emph{real} superpotential $P\equiv P(\varphi)$ and the scalar metric:
\begin{equation}\label{eq:VfromP_3dN1}
  V
  =
  \mathcal{G}^{MN}P_MP_N-4P^2\,,
  \qquad
  P_M \equiv \frac{\partial P}{\partial\varphi^M}\,,
\end{equation}
with $\mathcal{G}^{MN}$ the inverse of $\mathcal{G}_{MN}$.
In the following sections we will determine both $\mathcal{G}_{MN}$ and $P(\varphi)$ explicitly for the different string theories and orientifold choices considered in this paper.

%%%%%%%%%%%%%%%%%%%%%%%%%%%%%%%%%
%%%%%%%%%%%%%%%%%%%%%%%%%%%%%%%%%
%%%%%%%%%%%%%%%%%%%%%%%%%%%%%%%%%

\subsection{Three-dimensional form fields and dualizations}\label{ssec:vectordual}

In three dimensions, an abelian vector is dual to a scalar, so the same physical degree of freedom can be described either by the gauge field $A^n$ or equivalently by a dual scalar $\tilde c_n$. Starting from 
\begin{equation} 
(4\pi e)^{-1}\cL \supset -\frac{1}{4}f_{nm}(\varphi) F^n_{\mu\nu}F^{m\,\mu\nu}\,, \qquad F^n=\dd A^n\,, 
\end{equation} 
where the indices $n,m=1,\dots,N_{U(1)}\,,$ stand for the different $U(1)$ gauge fields, and $f_{nm}(\varphi)\equiv f_{nm}$ is the gauge kinetic matrix, which is generally a function of the moduli of the compactification.
The dualization can be implemented via the first-order formulation 
\begin{equation} 
(4\pi e)^{-1}\cL \supset -\frac{1}{4}f_{nm}(\varphi)F^n_{\mu\nu}F^{m\,\mu\nu} +\frac{1}{2}\epsilon^{\mu\nu\rho}\tilde c_n\partial_\mu F^n_{\nu\rho}\,. \label{eq:3dDualizationFirstOrder} 
\end{equation} 
Integrating out $\tilde c_n$ imposes the Bianchi identity for $F^n$, while eliminating $F^n$ gives 
\begin{equation} 
F^n = f^{nm}\star_3 \dd \tilde c_m\,, \qquad f^{nm}f_{mp}=\delta^n{}_p\,. 
\end{equation} 
Substituting back then yields the dual scalar action 
\begin{equation} 
(4\pi e)^{-1}\cL_{\mathrm{dual}} \supset -\frac{1}{2}\,f^{nm}\partial_\mu\tilde c_n \partial^\mu \tilde c_m\,. \label{eq:3dScalarFromVector} 
\end{equation}
We can equivalently introduce scalars with upper indices by defining
\begin{equation}
c^n \equiv \delta^{nm}\tilde c_m\, .
\end{equation}
Then the dual scalar action takes the form
\begin{equation}
(4\pi e)^{-1}\cL_{\mathrm{dual}} \supset -\mathcal{G}_{mn}\,\partial_\mu c^m \partial^\mu c^n\,,
\qquad
\mathcal{G}_{mn} = \frac12 (f^{-1})_{mn}
= \frac12 \delta_{mp}\delta_{nq} f^{pq}(\varphi)\,.
\label{eq:3dScalarFromVectorfinal}
\end{equation}

In three dimensions, a two-form potential has a three-form field strength,
which is a top form and therefore carries no propagating degrees of freedom.
It is dual to a constant flux parameter, and after integrating it out we
include the corresponding contribution in the scalar potential below.
Spacetime-filling three-form potentials are auxiliary in three dimensions.
Since they do not give rise to propagating fields in the effective theory,
we set them to zero and impose any associated tadpole
conditions separately.

\section{\texorpdfstring{Type IIA flux compactifications on $G_2$ orientifolds}{}}
\label{sec:IIA}

In this section we develop a generic three-dimensional effective description of type~IIA flux compactifications on $G_2$-holonomy manifolds, extending the analysis of \cite{Farakos:2020phe}.
Compactification of ten-dimensional type~IIA theory on a manifold with full $G_2$ holonomy yields $\mathcal{N}=2$ supersymmetry in three dimensions, corresponding to four real supercharges.
Introducing orientifold planes further projects the three-dimensional theory to $\mathcal{N}=1$, corresponding to two real supercharges.
We focus on the O2/O6 system, which is the standard supersymmetric option on genuine $G_2$ holonomy spaces \cite{Farakos:2020phe}, because as we discuss further below the alternative O4/O8 orientifold projection does not preserve supersymmetry unless the structure group is strictly smaller than $G_2$, an example of which was recently discussed in \cite{Tringas:2026ncg}.

\subsection{Ten-dimensional action and democratic RR sector}\label{subsec:IIA_10d_action}

In ten-dimensional Einstein frame the bosonic action of type~IIA supergravity can be written in the democratic formalism \cite{Bergshoeff:2001pv} as in equation \eqref{eq:S10_universal} with 
\begin{equation}
    \sum_\Lambda\frac{1}{4}e^{a_\Lambda\phi}\bigl|\mathcal{F}^{\Lambda}_{p_\Lambda}\bigr|^2 = \sum_{p\ \mathrm{even}}\frac{1}{4}\,e^{\frac{5-p}{2}\phi}\,|\tilde{F}_p|^2\,.
\end{equation}
In the democratic formalism one introduces all even-degree RR field strengths
$\tilde{F}_p$, with $p=0,2,\ldots,10$ and imposes the duality constraint
$\tilde{F}_p = (-1)^{\frac{p(p-1)}{2}}\star_{10}\tilde{F}_{10-p}$ at the level of the equations of motion. The Romans mass is $F_0$ and admits a dual description in terms of a 9-form potential / ten-form field strength \cite{Bergshoeff:2010mv}.

It is convenient to package RR fluxes in the polyform
\begin{equation}
  \mathbf{\tilde{F}}\equiv \sum_{p\ \mathrm{even}} \tilde{F}_p\,,
\end{equation}
whose Bianchi identity in the presence of localized sources reads
\begin{equation}\label{eq:RRpolyformBianchi}
  \dd \mathbf{\tilde{F}}
  =H_3\wedge \mathbf{\tilde{F}}
  +j_{\mathrm{loc}}\,.
\end{equation}
The $H_3$ flux satisfies $dH_3=0$.

We can write the above gauge-invariant polyform $\mathbf{\tilde F}$ in terms of a quantized background Page-flux polyform $\mathbf{F}^{\mathrm{bg}}\equiv \sum_{p\ \mathrm{even}} F_p^{\mathrm{bg}}$ and another polyform encoding the fluctuations $\mathbf{C} \equiv \sum_{p\ \mathrm{odd}} C_p$ via the equation
\be
\mathbf{\tilde{F}} = (\dd-H_3\w) \mathbf{C} + e^{B_2} \w \mathbf{F}^{\mathrm{bg}}\,.\label{eq:gaugeinvariantpolyform}
\ee

On a compact seven-manifold $X$ with $G_2$ holonomy, the zero modes expand as follows
\begin{align}
  \Phi &= s^I\Phi_I \,, \label{eq:IIA_phi_expand_unproj}\\
  B_2 &= b^A\Upsilon_A +B_{2}^{(3)}\,, \label{eq:IIA_B2_expand_unproj}\\
  C_3 &= c^I\Phi_I + A^A\wedge \Upsilon_A + C^{(3)}_{3}\,, \label{eq:IIA_C3_expand_unproj}
\end{align}
where $A^A$ are 3D abelian vectors and $B^{(3)}_{2}$, $C^{(3)}_{3}$ are the purely external 2- and 3-form components, respectively. (For $b^1(X)=0$, $C_1$ produces only a 3D vector.)

In 3D, all abelian vectors like the $A^A$ can be dualized to scalars; 2-forms and 3-forms carry no local degrees of freedom.
Hence, at the level of propagating bosonic degrees of freedom, the compactification can be described purely in terms of real scalars, which before the orientifold projection organize into $\mathcal{N}=2$ multiplets with four real supercharges.
The O-plane projection truncates this to $\cN=1$ and leaves only a subset of the above fields, depending on orientifold parities.

\subsection{The orientifold projection}
\label{subsec:IIA_O2O6_parities}

In type IIA, there are two possible orientifold projections, both realised as $\mathbb{Z}_2$ quotients involving the worldsheet parity operator, possibly combined with $(-1)^{F_L}$, and a spacetime involution.
The fixed loci of the spacetime involution correspond to spacetime-filling orientifold planes, whose dimensions must differ by four in order to preserve supersymmetry.
This leaves us with O4/O8-planes and O2/O6-planes.

Spacetime-filling O4/O8-planes are not compatible with a $G_2$-structure compactification, since in three dimensions they would wrap internal two- and six-dimensional cycles, which are not calibrated by the canonical $G_2$ forms \cite{HarveyLawson1982,Joyce2004}.
Equivalently, the corresponding orientifold involution is not among the standard supersymmetry-preserving $G_2$ involutions acting on the associative three-form as $\Phi \mapsto \pm \Phi$.
Thus, such sources would generically break the supersymmetry preserved by the $G_2$-structure, unless the structure group is further reduced, for instance to an $SU(3)$-structure.
Such cases, with O4/O6-planes, were studied recently in \cite{Tringas:2026ncg}, where the seven-dimensional geometry is described in terms of an $SU(3)$-structure six-manifold together with a circle, providing the additional calibration forms needed for the corresponding orientifold system.
We do not discuss these cases here, but restrict instead to genuine $G_2$-structure compactifications without a further reduction of the structure group.

The O2/O6 orientifolds arise from the involution
\begin{equation}
  \cO \;=\; \Omega_p\,(-1)^{F_L}\,\sigma\,,
  \label{eq:orientifoldOperator_O2O6}
\end{equation}
where $\sigma:X\to X$ is an involutive isometry of the target space, $\Omega_p$ is worldsheet parity, and $(-1)^{F_L}$ is the left-moving spacetime fermion number operator, acting as $-1$ on states with odd left-moving fermion number. The involution $\sigma$ is orientation reversing and acts on the $G_2$ structure as \cite{Farakos:2020phe}
\begin{equation}
  \sigma^*\Phi = -\,\Phi\,,
  \qquad
  \sigma^*\Psi = +\,\Psi\,,
  \label{eq:sigmaAction_phi_psi}
\end{equation}
so that the O6-plane fixed locus is a coassociative 4-cycle.
We will indicate this parity by superscripts and denote the forms $\Phi^-$ and $\Psi^+$.
This is the $G_2$ analogue of the anti-holomorphic involution in type~IIA Calabi--Yau O6 orientifolds \cite{Grimm:2004ua}. 

The involution $\sigma$ splits cohomology into even and odd eigenspaces,
\begin{equation}
  H^p(X,\R) = H^p_+(X)\oplus H^p_-(X)\,,
  \label{eq:Hsplit}
\end{equation}
and we choose parity-adapted bases
\begin{align}
  &\{\Upsilon_\alpha^+\}\subset H^2_+(X),\quad \alpha=1,\dots,b^2_+\,,
  &&\{\Upsilon_a^-\}\subset H^2_-(X),\quad a=1,\dots,b^2_-\,,
  \label{eq:bases2parity}\\
  &\{\Phi_i^+\}\subset H^3_+(X),\quad i=1,\dots,b^3_+\,,
  &&\{\Phi_\rho^-\}\subset H^3_-(X),\quad \rho=1,\dots,b^3_-\,.
  \label{eq:bases3parity}
\end{align}
Note that the top 7-form of the $G_2$ structure space is odd, so we have in this case $b_\pm^p = b_\mp^{7-p}$ and explicitly
\begin{align}
  &\{\Psi_i^-\}\subset H^4_-(X),\quad i=1,\dots,b^4_-\,,
  &&\{\Psi_\rho^+\}\subset H^4_+(X),\quad \rho=1,\dots,b^4_+\,,
  \label{eq:bases4parity}\\
  &\{T_\alpha^-\}\subset H^5_-(X),\quad \alpha=1,\dots,b^5_-\,,
  &&\{T_a^+\}\subset H^5_+(X),\quad a=1,\dots,b^5_+\,.
  \label{eq:bases5parity}
\end{align}
With \eqref{eq:sigmaAction_phi_psi}, the surviving metric moduli arise from the expansion in $\sigma$-odd forms
\begin{equation}
  \Phi^- = s^\rho\Phi_\rho^-\,.
  \label{eq:phiOddExpand}
\end{equation}

The parities of ten-dimensional fields under the spatial involution $\sigma$ (see \cite{Angelantonj:2002ct} for general orientifold
reviews and \cite{Farakos:2020phe} for the $G_2$ application) are
\begin{equation}\label{eq:fieldParities_O2O6}
  \{ g_{MN}\,,\,\phi\,,\,C_3 \} \quad \text{are}\quad \sigma \text{-even}\,,\qquad
  \{B_2\,,\,C_1\} \quad \text{are} \quad \sigma\text{-odd}\,.
\end{equation}
Hence the surviving Kaluza--Klein zero modes expand as
\begin{align}
  B_2 &= b^a\,\Upsilon_a^-\,, \label{eq:B2_expand_O2O6}\\
  C_3 &= c^i\,\Phi_i^+ + A^\alpha\wedge \Upsilon_\alpha^+\,,
  \label{eq:C3_expand_O2O6}
\end{align}
where the purely external two-form component $B_{2}^{(3)}$ is projected out.
The purely external component $C^{(3)}_3$ is a top-form potential in three dimensions and carries no local degrees of freedom, and is therefore omitted from the propagating spectrum.
Moreover, $C_1$ does not give rise to a three-dimensional vector field, since it is projected out by the orientifold action.

\subsection{The kinetic terms}
\label{subsec:IIA_kinetic_terms_generic}

We now collect the kinetic terms for the propagating bosons after the O2/O6 projection and, when present, after dualizing the surviving 3D vectors.
Restricting the geometric overlap metrics of Section~\ref{sec:G2prelim} to the orientifold parity eigenspaces, we define
\begin{align}
  G^-_{\rho\sigma}
  &= \frac{1}{\cV}\int_X \Phi^-_\rho \wedge \star \Phi^-_\sigma\,,
  &
  G^-_{ab}
  &= \frac{1}{\cV}\int_X \Upsilon^-_a \wedge \star \Upsilon^-_b\,,
  \label{eq:IIA_metrics_minus}\\
  G^+_{ij}
  &= \frac{1}{\cV}\int_X \Phi^+_i \wedge \star \Phi^+_j\,,
  &
  G^+_{\alpha\beta}
  &=\frac{1}{\cV}\int_X \Upsilon^+_\alpha \wedge \star\Upsilon^+_\beta\,,
  \label{eq:IIA_metrics_plus}
\end{align}
and similarly for the related scalar field metrics $\mathcal{G}_{MN}$ from Section \ref{sec:10dto3d_general}.
All these metrics have $s$-moduli dependence.

The universal dilaton kinetic term and the kinetic term for the surviving odd $G_2$ moduli were derived above in subsection \ref{subsec:3d_Einstein_frame} and are given by
\begin{equation}\label{eq:xyKinetic}
  (4\pi e)^{-1}\cL_{\phi,s}
  = -\frac14\,\partial_\mu\phi\partial^\mu\phi
    -\mathcal{G}^-_{\rho\sigma}\partial_\mu s^\rho\partial^\mu s^\sigma \,.
\end{equation}

Let us now consider the kinetic terms arising for the axions. For the NSNS axions $b^a$, expanded as in \eqref{eq:B2_expand_O2O6}, the external-derivative term yields
\begin{equation}\label{eq:Kb_axions}
  (4\pi e)^{-1}\cL_b
  = -\mathcal{G}^-_{ab}\,\partial_\mu b^a\partial^\mu b^b\,,\quad\quad\text{with}\quad\quad 
  \mathcal{G}^-_{ab}=\frac14\,e^{-\phi}G^-_{ab}\,,
\end{equation}
which is the truncation of the general result derived in Section~\ref{sec:10dto3d_general}, see equation~\eqref{eq:kineticmetricb_generic}.

We now turn to the axions arising in the RR-sector from the expansion \eqref{eq:C3_expand_O2O6}. We decompose the exterior derivative as $\dd = \dd_3 + \dd_7$, where $\dd_3$ acts on the external spacetime and $\dd_7$ on the internal seven-manifold. There are two relevant contributions that are external 1-forms
\begin{equation}
\dd C_3 \supset \dd_3 c^i \w\Phi_i^+ - A^\alpha \wedge \dd_7\Upsilon_\alpha^+ = (\dd_3 c^i - A^\alpha N_\alpha{}^i) \w\Phi_i^+ \equiv (D c^i) \w \Phi_i^+\,, \label{eq:gaugecovariant}
\end{equation} 
where the matrix $N$ was defined above in \eqref{eq:dUpsilonMatrix}. So, we see that for $G_2$ structure spaces the axions can carry a charge under the $U(1)$ gauge groups arising from the $A^\alpha$. Including the dilaton factor in   $e^{\phi/2}|\widetilde{F}_4|^2$, we obtain the following gauge covariant kinetic term
\begin{equation}
  (4\pi e)^{-1}\cL_c
  = -\mathcal{G}^+_{ij}D_\mu c^iD^\mu c^j\,,
  \quad\quad\text{with}\quad\quad 
  \mathcal{G}^+_{ij}=\frac14\,e^{\phi/2}G^+_{ij}\,.
  \label{eq:Kc_axions}
\end{equation}
This is the specialization of the general formula obtained above in equation \eqref{eq:kineticmetrica_generic}.

Note that the geometric flux induces gauging of axionic shift symmetries through the St\"uckelberg-type couplings $A^\alpha\wedge N_\alpha{}^i\,\Phi^+_i$ appearing in \eqref{eq:gaugecovariant}.
That means when $N_\alpha{}^i \neq 0$, the vectors $A^\alpha$ acquire a St\"uckelberg mass by gauging the shift symmetry of the RR scalars $c^i$.
In three dimensions, a massive vector carries two degrees of freedom and cannot be trivially dualized to a single free scalar $c^\alpha$ as was described above in subsection \ref{ssec:vectordual} for the massless case.
Consequently, for directions where $N_\alpha{}^i \neq 0$, the continuous shift symmetry is gauged and the effective theory retains massive vector multiplets.
The flux superpotential below accommodates this gauging: under a shift $\delta c^i = \lambda^\alpha N_\alpha{}^i$, the variation of the superpotential given in equation \eqref{eq:PIIA_O2O6_phiV} below is proportional to $N_\alpha{}^i M_i{}^\rho $, where the matrix $M$ is defined in \eqref{eq:dPhiMatrix}.
Due to the geometric constraint  \eqref{eq:d2constraints}, this identically vanishes, rendering the flux superpotential exactly gauge-invariant.

The preceding discussion also clarifies which vectors can be dualized. The condition is that the corresponding linear combination of vectors lies in the kernel of the gauging map. Let
\[
  k_{\hat\alpha}{}^\alpha N_\alpha{}^i=0,
  \qquad
  \hat\alpha=1,\ldots,\dim\ker N ,
\]
be a basis for this kernel.  Completing it to a basis of the vector space,
we write
\[
  A^\alpha
  =
  k_{\hat\alpha}{}^\alpha A^{\hat\alpha}
  +
  \ell_{\bar\alpha}{}^\alpha A^{\bar\alpha}.
\]
In this basis the covariant derivatives take the form
\[
  D c^i
  =
  \dd_3 c^i - N_{\bar\alpha}{}^i A^{\bar\alpha},
  \qquad
  N_{\bar\alpha}{}^i\equiv \ell_{\bar\alpha}{}^\alpha N_\alpha{}^i ,
\]
while the kernel vectors $A^{\hat\alpha}$ do not gauge the RR axions. Therefore only the ungauged vectors $A^{\hat\alpha}$ can be dualized to scalars in the standard three-dimensional way.\footnote{Since $\widetilde F_4 = \star_{10} \widetilde F_6$, we could also get these dual axionic scalars directly from expanding $C_5$ in terms of the $\{T_{\hat\alpha}^-\}\subset H^5_-(X)$.}  The complementary vectors $A^{\bar\alpha}$ participate in St\"uckelberg couplings and should be kept as massive vectors.

Restricting now to the ungauged sector, the vectors arise from
\[
  \dd C_3 \supset
  \dd_3 A^{\hat\alpha}\wedge
  \Upsilon^+_{\hat\alpha}
  \equiv
  F^{\hat\alpha}\wedge \Upsilon^+_{\hat\alpha},
  \qquad
  \Upsilon^+_{\hat\alpha}
  \equiv
  k_{\hat\alpha}{}^\alpha\Upsilon^+_\alpha .
\]
Dimensional reduction of the term $e^{\phi/2}|\widetilde F_4|^2$ gives
\begin{equation}
  (4\pi e)^{-1}\,\cL_A
  =
  -\frac14\,f_{\hat\alpha\hat\beta}
  F^{\hat\alpha}_{\mu\nu}F^{\hat\beta\,\mu\nu}\, ,
  \label{eq:gaugeKineticMatrix}
\end{equation}
where
\begin{equation}
  f_{\hat\alpha\hat\beta}
  =
  \frac12\,e^{\phi/2}\,\cV
  \int_X \Upsilon^+_{\hat\alpha}\wedge \star \Upsilon^+_{\hat\beta}
  =
  k_{\hat\alpha}{}^\alpha k_{\hat\beta}{}^\beta
  f_{\alpha\beta}.
\end{equation}
Dualizing the vectors according to \eqref{eq:3dScalarFromVector} gives
\begin{equation}\label{eq:Kdual_axions}
  (4\pi e)^{-1}\,\cL_{\tilde c}
  =
  -\frac12\,f^{\hat\alpha\hat\beta}
  \partial_\mu\tilde c_{\hat\alpha}
  \partial^\mu\tilde c_{\hat\beta}\, .
\end{equation}
We can now introduce scalars with upper indices by defining $c^{\hat\alpha} \equiv \delta^{\hat \alpha\hat \beta}\tilde c_{\hat \beta}$. Then the dual scalar action takes the form \begin{equation}\label{eq:3dScalarFromVectorfinalIIA} 
(4\pi e)^{-1}\,\cL_{c} = -\mathcal{G}_{\hat\alpha\hat\beta}\partial_\mu c^{\hat\alpha} \partial^\mu c^{\hat \beta}\,, \qquad \mathcal{G}_{\hat\alpha\hat\beta}= e^{-\phi/2}\cV^{-2} \delta_{\hat\alpha\hat\gamma}\delta_{\hat\beta\hat\delta}(G_+^{-1})^{\hat\gamma\hat\delta}\,,
\end{equation}
where $(G_+^{-1})^{\hat\gamma\hat\delta}$ denotes the matrix inverse of the projected geometric metric $G^+_{\hat\alpha\hat\beta} \equiv k_{\hat\alpha}{}^\alpha k_{\hat\beta}{}^\beta G^+_{\alpha\beta}$.

Collecting the propagating bosons, the scalar fields can be taken to be
\begin{equation}
  \varphi^M
  =
  \{\phi,\ s^\rho,\ b^a,\ c^i,\ c^{\hat\alpha}\}\, .
  \label{eq:scalarSet_O2O6}
\end{equation}
The sigma-model metric is block diagonal in these coordinates, with blocks
given by \eqref{eq:xyKinetic}, \eqref{eq:Kb_axions}, \eqref{eq:Kc_axions}, and \eqref{eq:3dScalarFromVectorfinalIIA}. Note that the three-dimensional effective theory additionally contains the massive propagating vector fields $A^{\bar\alpha}$.

\subsection{Expanding the field strengths}
\label{subsec:IIA_torsion_parity}

On a $G_2$-structure space used in a consistent truncation, the truncated internal basis forms need not be closed.
However, since the orientifold involution $\sigma$ commutes with $\dd=\dd_3 +\dd_7$, the latter preserves the $\pm$ parity decomposition of the forms above.
Therefore, the internal derivatives can lead to extra axionic terms in the gauge-invariant field strengths from $\dd_7 B_2 = b^a\dd_7\Upsilon^-_a$, and $\dd_7 C_3 \supset c^i\dd_7\Phi^+_i$.
These arise in addition to the usual background parts that involve quantized fluxes and axions from the $e^{B_2}$ factor in equation \eqref{eq:gaugeinvariantpolyform}.
The gauge-invariant field strengths are explicitly given by
\begin{align}
  H_3 &= H_3^\mathrm{bg} + \dd B_2
   = \dd_3 b^a\wedge \Upsilon^-_a
     + h^\rho \Phi_\rho^-  + b^a \dd_7\Upsilon_a^-  \equiv \dd_3 b^a\wedge \Upsilon^-_a
     + H_3^\mathrm{int}\,,
  \label{eq:H3hat_parity_split}\\
  \tilde F_2
  &= F_2^\mathrm{bg} + F_0B_2
   = \big(e^a + F_0 \,b^a\big)\Upsilon^-_a \equiv \tilde F_2^\mathrm{int}\,,
  \label{eq:F2tilde_parity_split}\\
  \tilde F_4
  &= F_4^\mathrm{bg} + \dd C_3 + B_2\wedge F_2^\mathrm{bg}
      + \frac12 F_0\,B_2\wedge B_2 \notag\\
  &= \dd_3 c^i \wedge \Phi_i^+
   + F^\alpha \wedge \Upsilon_\alpha^+
   - A^\alpha \wedge \dd_7 \Upsilon_\alpha^+
   + f^\rho \Psi_\rho^+
   + c^i \dd_7 \Phi_i^+
   + B_2\wedge F_2^\mathrm{bg}
   + \frac12 F_0\,B_2\wedge B_2 \notag\\
  &\equiv
   (\dd_3 c^i - A^\alpha N_\alpha{}^i )  \wedge \Phi_i^+
   + F^\alpha \wedge \Upsilon_\alpha^+
   + \tilde F_4^\mathrm{int},
  \label{eq:F4hat_parity_split}
\end{align}
where we have defined internal parts of the field strengths that will be important for the explicit reduction below. 
The expansion coefficients $(h^\rho,F_0,e^a,f^\rho)$ that arose from expanding $(H_3^\mathrm{bg}, F_0, F_2^\mathrm{bg}, F_4^\mathrm{bg})$ denote the quantized background flux data. 

The relevant Bianchi identities from equation \eqref{eq:RRpolyformBianchi} above are
\begin{align}
  \dd H_3 & =0\,, \label{eq:BI_H3}\\
  \dd \tilde{F}_2 &= H_3F_0 + j_{\mathrm{O6}} + j_{\mathrm{D6}} \,, \label{eq:BI_O6}\\
  \dd \tilde{F}_4 &= H_3\w \tilde{F}_2\,,\label{eq:BI_O4inO2O6} \\
  \dd \tilde{F}_6 &= H_3\wedge \tilde{F}_4 + j_{\mathrm{O2}} + j_{\mathrm{D2}} \,,\label{eq:BI_O2}
\end{align}
where $j_{\mathrm{Op}}$ and $j_{\mathrm{Dp}}$ are the Poincar\'e-dual source currents, including the appropriate charge factors. These relations show which background flux combinations are constrained by tadpole cancellation:
\begin{itemize}
  \item The O6 tadpole \eqref{eq:BI_O6} constrains the cohomology class of $\dd \tilde{F}_2 - H_3 F_0$ in terms of the net O6/D6 charge.
  \item The O2 tadpole \eqref{eq:BI_O2} constrains only the component of $F_4$ that has non-trivial pairing with $H_3$.  Any component of $F_4$ in the kernel of $[F_4]\mapsto [H_3]\wedge[F_4]$ is not fixed by the O2/D2 tadpole.
\end{itemize}
In the presence of torsion/geometric fluxes the Bianchi identities constrain the allowed flux quanta. For example, using equation \eqref{eq:dPhiMatrix} we find from $\dd H_3 =0$ that $h^\rho M_{\rho}{}^{i}=0$.

\subsection{Real superpotential and scalar potential}
\label{subsec:IIA_superpotential_O2O6}

After dimensional reduction, the scalar potential $V$ of the effective theory is
\begin{align}
  V
  &=
  V_{\rm curv}+V_{\rm flux}+V_{\rm loc}\,,
  \label{eq:Vflux_O2O6_phiV}\\[0.3em]
  V_{\rm curv}
  &=
  -\frac{1}{2\mathcal{V}^3}
  \int_X d^7y\,\sqrt{g_7}\,R_7 \,,\\
  V_{\rm flux}
  &=
  \frac{1}{4\mathcal{V}^3}
  \int_X
  \Big(
  e^{-\phi} H_3^\mathrm{int}\wedge\star H_3^\mathrm{int}
  + e^{5\phi/2} F_0\wedge\star F_0\cr
  &\qquad\qquad\quad 
  + e^{3\phi/2} \widetilde F_2^\mathrm{int}\wedge\star \widetilde F_2^\mathrm{int}
  + e^{\phi/2} \widetilde F_4^\mathrm{int}\wedge\star \widetilde F_4^\mathrm{int}
    \Big)\,,\\
  V_{\rm loc}
  &=
  V_{O2/D2}+V_{O6/D6}\,.
\end{align}
The real superpotential that reproduces \eqref{eq:Vflux_O2O6_phiV} through the standard three-dimensional formula \eqref{eq:VfromP_3dN1} is
\begin{equation}\label{eq:PIIA_O2O6_phiV}
  P_{\rm IIA}
  =
  \frac{1}{4\cV^2}
  \int_X
  \left(
    e^{-\phi/2}\,\Psi\wedge H_3^\mathrm{int}
    + e^{\phi/4}\,\Phi\wedge \widetilde F_4^\mathrm{int}
    + e^{5\phi/4}\,\star F_0
  \right) \,. 
\end{equation}
While this superpotential appears nearly identical to one previously presented in the literature, see \cite{Farakos:2020phe}, it contains several additional terms that are hidden by the compact notation, as we explain below.
For the sake of compactness, we drop the subscript ``$7$" from $\dd_7$ and call it just $\dd$ for the rest of the section.  

One of the main extensions relative to previous Ricci-flat massive IIA setups is the inclusion of curvature, which appears here in a non-trivial way, as we will see.
In contrast to type IIB/type I compactifications, where the curvature contribution arises in the superpotential from a term of the form $\int_X \Phi \wedge \dd \Phi$, such a term is absent in massive IIA on $G_2$ compactifications.
The reason is that $\Phi \wedge \dd \Phi$ is an even seven-form, whereas the top form in the present setup is odd.
Nevertheless, the curvature contribution $V_{\rm curv}$ to the scalar potential is encoded in the superpotential \eqref{eq:PIIA_O2O6_phiV}, and in particular receives two contributions that are hidden in this expression.
The first is in $H_3^\mathrm{int} \supset b^a \dd \Upsilon_a^-$ from equation \eqref{eq:H3hat_parity_split} above and this leads to 
\be
\mathcal{G}^{ab} \partial_{b^a} P_{\rm IIA} \partial_{b^b} P_{\rm IIA} \supset \frac{1}{4 \cV^3} \int_X \dd \Psi^+ \w \star \dd \Psi^+ 
=\frac{1}{4 \cV^2} |\tau_2|^2\,.
\ee
The second contribution is in $\tilde{F}_4^\mathrm{int} \supset c^i \dd \Phi_i^+$ from equation \eqref{eq:F4hat_parity_split} above and leads to 
\be
\mathcal{G}^{ij} \partial_{c^i} P_{\rm IIA} \partial_{c^j} P_{\rm IIA} \supset \frac{1}{4 \cV^3} \int_X \dd \Phi^- \w \star \dd \Phi^- 
=\frac{1}{4 \cV^2} |\tau_3|^2\,.
\ee

From equation \eqref{eq:dPhiTorsion} with the added orientifold parities in equation \eqref{eq:sigmaAction_phi_psi} as superscripts, we find that in our case
\begin{equation}
  \dd \Phi^- = \tau_0\,\Psi^+ + 3\,\tau_1\wedge \Phi^- + \star \tau_3 \,.
\end{equation}
We are working with smeared sources, constant fluxes, and torsion classes, so that $\tau_0$ would have to be an odd constant, which means $\tau_0=0$.
Furthermore, since there are no 1-forms on our compact $G_2$ structure space, we have that $\tau_1=0$ as well.
This means that the curvature in equation \eqref{eq:R7viaTorsion} reduces to $R_7 =- \frac12 |\tau_2|^2 -\frac12 |\tau_3|^2$. Thus,
\begin{equation}
   V_{\text{curv}}=-\frac{1}{2\mathcal{V}^3}\int_Xd^7y \sqrt{g_7}R_7 = \frac{1}{4\mathcal{V}^2}(|\tau_2|^2+|\tau_3|^2)\,,
\end{equation}
is exactly reproduced by the two terms above. Note also that this means that in type IIA on $G_2$ structure spaces we have always a non-positive curvature $R_7\leq0$ and therefore $V_{\text{curv}}\geq 0$.

Another interesting feature of \eqref{eq:PIIA_O2O6_phiV} is that no additional independent linear term in $\widetilde F_2^\mathrm{int}$ is needed. The term $\int_X \Phi\wedge B_2\wedge\widetilde F_2^\mathrm{int}$ is already contained in $\Phi\wedge\widetilde F_4^\mathrm{int}$, with exactly the right prefactor. We have
\begin{equation}\label{eq:Pa_from_F2}
  \partial_{b^a}P_{\rm IIA}
  \supset
  \frac{e^{\phi/4}}{4\cV^2}
  \int_X \Phi\wedge \Upsilon^-_a\wedge \widetilde F_2^\mathrm{int}
  =
  -\frac{e^{\phi/4}}{4\cV}
  G^-_{ab}(e^b+F_0 b^b)\,,
\end{equation}
where we used \eqref{eq:starUpsilon}.
With the $B_2$-axion metric $\cG_{ab}=\frac14 e^{-\phi}G^-_{ab}$ from Section~\ref{subsec:IIA_kinetic_terms_generic}, this gives
\begin{equation}
  \cG^{ab}P_aP_b
  \supset
  \frac{1}{4\cV^3}
  \int_X e^{3\phi/2}\widetilde F_2^\mathrm{int}\wedge\star \widetilde F_2^\mathrm{int}\,,
\end{equation}
precisely matching the $\widetilde F_2^\mathrm{int}$ contribution in \eqref{eq:Vflux_O2O6_phiV}. 
The same overall prefactor $1/(4\cV^2)$ in \eqref{eq:PIIA_O2O6_phiV} is also the one required for the $H_3^\mathrm{int}$, $\widetilde F_4^\mathrm{int}$, and Romans-mass terms.

Finally, the O2-plane and O6-plane contributions are not introduced as separate terms in the superpotential. 
Rather, their contributions to the scalar potential are reproduced through cross terms between the flux contributions, in particular those involving the Romans mass or the $\tilde F_4^\mathrm{int}$ sector together with $H_3^\mathrm{int}$, after imposing the Bianchi identities in \eqref{eq:BI_O2} and \eqref{eq:BI_O6}, as shown in \cite{Farakos:2020phe}.

Thus, \eqref{eq:PIIA_O2O6_phiV} is the generalized superpotential in the presence of $F_2$ flux, curvature and axions, and it precisely reproduces the dimensionally reduced 10d action for a generic compactification on a $G_2$ structure space.

\section{\texorpdfstring{Type IIB/I flux compactifications on $G_2$ orientifolds}{}}
\label{sec:IIB}

In this section we develop the three-dimensional effective description of supersymmetric type~IIB orientifold flux compactifications on $G_2$ manifolds and, more generally, on $G_2$-structure spaces with a consistent truncation.
Our discussion parallels the type~IIA reduction of \cite{Farakos:2020phe} and generalizes the type~IIB/I analysis of \cite{Emelin:2021gzx}; for the closely related Calabi--Yau orientifold case see \cite{GrimmLouisIIB0403067}.
We focus on supersymmetric setups with O5-planes and then explain how the O9 projection truncates them to the type~I closed-string sector.

\subsection{Ten-dimensional action and democratic RR sector}
\label{subsec:IIB10d}

In ten-dimensional Einstein frame the bosonic type~IIB pseudo-action can be written in the democratic formalism \cite{Bergshoeff:2001pv} as in equation~\eqref{eq:S10_universal} with
\begin{equation}
  \sum_\Lambda \frac14 e^{a_\Lambda \phi} \bigl|\mathcal F^{\Lambda}_{p_\Lambda}\bigr|^2
  =
  \sum_{p\ \mathrm{odd}} \frac14\, e^{\frac{5-p}{2}\phi}\, |\widetilde F_p|^2\,.
\end{equation}
In the democratic formalism one introduces all odd-degree RR field strengths $\widetilde F_p$, with $p=1,3,5,7,9$, and imposes the duality constraint 
\begin{equation}
  \widetilde F_p = (-1)^{\frac{p(p-1)}{2}} \star_{10} \widetilde F_{10-p}\label{eq:FpIIBduality}
\end{equation}
at the level of the equations of motion~\cite{Bergshoeff:2001pv}.

As in the type~IIA discussion, it is convenient to organize the RR fluxes into the odd polyform
\begin{equation}
  \widetilde{\mathbf F} \equiv \sum_{p\ \mathrm{odd}} \widetilde F_p \,,
\end{equation}
with Bianchi identity
\begin{equation}\label{eq:IIB_RRpolyformBianchi}
  \dd \widetilde{\mathbf F}
  =
  H_3 \wedge \widetilde{\mathbf F}
  + j_{\mathrm{loc}} \,,
\end{equation}
while the $H_3$ flux satisfies $dH_3=0$.
We can write the above polyform in terms of a quantized background flux polyform $\mathbf{F}^{\mathrm{bg}}\equiv \sum_{p\ \mathrm{odd}} F_p^{\mathrm{bg}}$ and a polyform encoding the fluctuations $\mathbf{C} \equiv \sum_{p\ \mathrm{even}} C_p$ via the equation
\begin{equation}
\mathbf{\tilde{F}} = (\dd-H_3\w) \mathbf{C} + e^{B_2} \w \mathbf{F}^{\mathrm{bg}}\,.\label{eq:tildeFIIB}
\end{equation}

On a compact seven-manifold $X$ with $G_2$ holonomy, and hence $b^1(X)=0$, the unprojected zero modes descend as
\begin{align}
  \Phi &= s^I \Phi_I\,, \label{eq:IIB_phi_expand_unproj}\\
  B_2 &= b^A \Upsilon_A\,, \label{eq:IIB_B2_expand_unproj}\\
  C_2 &= c_2^A \Upsilon_A\,, \label{eq:IIB_C2_expand_unproj}\\
  C_4 &= c_4^I \Psi_I + A^I \wedge \Phi_I + B^{A}_{(2)} \wedge \Upsilon_A\,.
  \label{eq:IIB_C4_expand_unproj}
\end{align}
Here $A^I$ are three-dimensional abelian vectors and $B^{A}_{(2)}$ are three-dimensional 2-forms.
The two-forms and any external 3-forms are non-dynamical in three dimensions.
As in the type~IIA case, the propagating bosonic degrees of freedom can be described entirely in terms of scalars after dualizing the abelian 3D vectors.
However, in type IIB we have to impose the self-duality of $\widetilde F_5$, see \eqref{eq:FpIIBduality}, which removes half of its dynamical components.
In this case we can therefore simply work with the $c_4^I$ axions since they are the 3D scalars dual to the gauge fields $A^I$.

\subsection{The orientifold projection}
\label{subsec:IIB_O5O9_parities}

In type~IIB there are two possible classes of spacetime-filling orientifold planes whose dimensions differ by four, namely O5/O9 and O3/O7 \cite{Angelantonj:2002ct,GrimmLouisIIB0403067}.
On a strict $G_2$ background, however, only the O5/O9 class is compatible with the single real internal spinor defining the $G_2$ structure.
Geometrically, in a compactification to three dimensions an O5-plane wraps an internal three-cycle, which can be associative and hence calibrated by $\Phi$, while an O9-plane fills the internal seven-manifold.
By contrast, space-filling O3- and O7-planes would wrap internal one- and five-dimensional cycles, respectively, which are not calibrated by the canonical $G_2$ calibration forms $\Phi$ and $\Psi$ \cite{HarveyLawson1982,Joyce2004}.

The same conclusion follows from the spinorial lift of the orientifold involution. If the involution $\sigma$ reflects $r=9-p$ internal coordinates transverse to an Op-plane, then locally its lift to the internal spin bundle is, up to an overall sign,
\[
  \widehat{\sigma}=\Gamma_{a_1}\cdots \Gamma_{a_r},
  \qquad
  \widehat{\sigma}^{\,2}=(-1)^{r(r-1)/2}.
\]
For O9- and O5-planes one has $r=0$ and $r=4$, respectively, so $\widehat{\sigma}^{\,2}=+1$ and the unique real $G_2$ spinor can be assigned a definite orientifold parity. For O7- and O3-planes one has $r=2$ and $r=6$, respectively, so $\widehat{\sigma}^{\,2}=-1$; equivalently, the lift has eigenvalues $\pm i$ after complexification. A strict $G_2$ compactification has only one real internal Majorana spinor, so there is no real invariant spinor satisfying the O3/O7 projection. Realising the O3/O7 branch would require at least two real internal spinors, i.e. a reduction of the structure group below $G_2$. We therefore restrict to the O5/O9 orientifold class.

The O5/O9 orientifolds arise from the involution
\begin{equation}
  \cO = \Omega_p\sigma\,,
  \label{eq:IIB_O5O9_operator}
\end{equation}
where $\sigma:X\to X$ is an involutive isometry~\cite{Angelantonj:2002ct}.
Both O5- and O9-planes are associated with an orientation-preserving internal involution, under which supersymmetry requires
\begin{equation}
  \sigma^* \Phi = +\Phi \,,
  \qquad
  \sigma^* \Psi = +\Psi \,.
  \label{eq:IIB_sigmaAction_phi_psi_O5O9}
\end{equation}
Thus the top form is even, and the Hodge star preserves orientifold parity.

The involution splits cohomology as in \eqref{eq:Hsplit}, and we keep the same index conventions as in \eqref{eq:bases2parity}--\eqref{eq:bases3parity}, namely
\begin{align}\label{eq:IIB_bases23parity}
  &\{\Upsilon_\alpha^+\}\subset H^2_+(X),\quad \alpha=1,\dots,b^2_+\,,
  &&\{\Upsilon_a^-\}\subset H^2_-(X),\quad a=1,\dots,b^2_-\,,
  \\
  &\{\Phi_i^+\}\subset H^3_+(X),\quad i=1,\dots,b^3_+\,,
  &&\{\Phi_\rho^-\}\subset H^3_-(X),\quad \rho=1,\dots,b^3_-\,.
\end{align}
Because the top form is now even, the Hodge-dual parity assignments are
\begin{align}\label{eq:IIB_bases45parity}
  &\{\Psi_i^+\}\subset H^4_+(X),\quad i=1,\dots,b^3_+\,,
  &&\{\Psi_\rho^-\}\subset H^4_-(X),\quad \rho=1,\dots,b^3_-\,,\\
  &\{T_\alpha^+\}\subset H^5_+(X),\quad \alpha=1,\dots,b^2_+\,,
  &&\{T_a^-\}\subset H^5_-(X),\quad a=1,\dots,b^2_-\,.
\end{align}
With \eqref{eq:IIB_sigmaAction_phi_psi_O5O9}, the surviving metric moduli arise from the expansion in $\sigma$-even forms
\begin{equation}
  \Phi^+ = s^i \Phi_i^+\,.
  \label{eq:IIB_phiEvenExpand}
\end{equation}

The parities of the ten-dimensional bosonic fields under $\sigma$, following the standard IIB orientifold conventions \cite{Angelantonj:2002ct,GrimmLouisIIB0403067}, are
\begin{equation}\label{eq:IIB_O5O9_field_parities}
  \{g_{MN},\,\phi,\,C_2,\,C_6\}
  \quad \text{are} \quad \sigma\text{-even}\,,
  \qquad
  \{B_2,\,C_0,\,C_4\}
  \quad \text{are} \quad \sigma\text{-odd}\,.
\end{equation}
Hence the surviving light closed-string fields in the pure O5-plane branch expand as
\begin{equation}\label{eq:IIB_C4_expand_O5}
  B_2 = b^a\,\Upsilon_a^-\,,
  \quad\quad
  C_2 = c_2^\alpha\,\Upsilon_\alpha^+\,,
  \quad\quad
  C_4 = c_4^\rho\,\Psi_\rho^-\,,
\end{equation}
where we have dropped the vector fields $A^\rho$, that are related to the axions $c_4^\rho$ through the self-duality of $\widetilde{F}_5$, as well as the non-dynamical three dimensional 2-forms.
Since $b^0_-(X)=0$ on a connected compact manifold, $C_0$ is projected out entirely.

If an O9 projection is also imposed, one must further keep only the $\Omega_p$-even subsector.
Since $B_2$, $C_0$, and $C_4$ are odd under $\Omega_p$, the O9 projection sets
\begin{equation}
  B_2 = 0\,,
  \qquad
  C_0 = 0\,,
  \qquad
  C_4 = 0\,,
  \qquad
  \widetilde F_5 = 0\,.
  \label{eq:IIB_O9_truncation}
\end{equation}
The remaining closed sector is then precisely the strict type~I truncation of the O5-plane branch.

\subsection{The kinetic terms}
\label{subsec:IIB_O5O9_kinetic}

Restricting the geometric overlap metrics of Section~\ref{sec:G2prelim} to the orientifold parity eigen-spaces, leads to the same matrices as above in equation \eqref{eq:IIA_metrics_minus} and \eqref{eq:IIA_metrics_plus}.
It is also useful to define 
\begin{align}
    \hat{G}^{-}_{\rho \sigma} = \frac{1}{\mathcal{V}} \int_X \Psi_{\rho}^{-} \wedge \star \Psi_\sigma^{-} \label{eq:IIB_c4metric}
\end{align}
which will appear below in the kinetic term for the scalars $c_4^\rho$ retained in the $C_4$ truncation.
As in the type~IIA reduction, all these metrics depend on the geometric moduli~$s^i$.

The universal dilaton kinetic term and the kinetic term for the surviving even $G_2$ moduli are
\begin{equation}
  (4\pi e)^{-1}\cL_{\phi,s}
  = -\frac14\,\partial_\mu\phi\,\partial^\mu\phi
    -\mathcal G^+_{ij}\,\partial_\mu s^i\,\partial^\mu s^j\,.
  \label{eq:IIB_kin_phi_s}
\end{equation}
For the RR axions $c_2^\alpha$ descending from $C_2$, one finds
\begin{equation}
  (4\pi e)^{-1}\cL_{c_2}
  = -\mathcal G^+_{\alpha \beta} \partial_\mu c_2^{\alpha} \partial^\mu c_2^{\beta}\,,
  \qquad
  \mathcal G^+_{\alpha \beta}=\frac14\,e^{\phi} G^+_{\alpha \beta}\,.
  \label{eq:IIB_kin_b}
\end{equation}

Now we turn to the kinetic terms of the RR axions $c_4^\sigma$ and $b^a$. From equation \eqref{eq:tildeFIIB}, $\tilde{F}_5$ contains the following terms
\begin{equation}
\begin{split}
    \tilde{F}_5 &\supset \dd C_4 -H_3 \w C_2 \\
    &\supset \dd_3c_4^\rho\ \Psi_\rho^- -\dd_3 b^a\,\Upsilon_a^- \w c_2^\alpha\,\Upsilon_\alpha^+ \,.
\end{split}
\end{equation}
Thus from the $|\tilde{F}_5|^2 $ term, there is a kinetic mixing between the $c_4^\rho$ and $b^a$ axions that depends on the axions in $C_2$. Additionally, we get from the $|H_3|^2$ term the same kinetic term for the $b^a$ axions as in type IIA above in equation \eqref{eq:Kb_axions}.

It is convenient to express the kinetic terms of the combined sector, $c^m = (b^a, c_4^\rho )$, in the following way,
\begin{equation}\label{eq:IIB_kin_c}
  (4\pi e)^{-1}\cL_c
  = -\mathcal G_{m n}\partial_\mu c^m\partial^\mu c^n\,.
\end{equation}
The metric $\mathcal{G}_{m n}$ written in matrix form is,
\begin{align}
\mathcal{G}
& =
\frac14
\begin{pmatrix}
e^{-\phi} G^- + S^{T} \hat{G}^- S &{} & {} & {}&   -S^{T}\hat{G}^-\\[2mm]
-\hat{G}^- S & {} & {} & {}& \hat{G}^-
\end{pmatrix}\,,
\label{eq:IIB_cmetric}
\end{align} 
where the matrix $S(c_2^\alpha)$ depends on the $C_2$ axions and is defined through the expansion
\begin{equation}
  \Upsilon_a^- \w C_2 
  = \Upsilon_a^- \w c_2^\alpha\Upsilon_\alpha^+ = 
  S^\rho{}_a\Psi_\rho^- \,,
\end{equation}
while $G^-$ and $\hat{G}^-$ are the metrics introduced in \eqref{eq:IIA_metrics_minus} and \eqref{eq:IIB_c4metric}, respectively.

The set of real scalar coordinates is therefore
\begin{equation}
  \varphi^M
  = \{\phi,\ s^i,\ b^a,\ c_2^\alpha,\ c_4^\rho\}\,.
  \label{eq:IIB_scalar_set}
\end{equation}

\subsection{Expanding the field strengths}
\label{subsec:IIB_O5O9_fluxes}

On a $G_2$-structure space used in a consistent truncation, the truncated internal basis forms need not be closed \cite{DallAgata:2005zlf, Emelin:2021gzx}.
We again decompose $\dd = \dd_3 + \dd_7$, where $\dd_3$ acts on the external spacetime and $\dd_7$ on the internal seven-manifold.
Since $\sigma$ commutes with $\dd_7$, the latter preserves orientifold parity.
As above in the type IIA setup, the non-closure of the internal basis forms will lead to new axion dependencies that have not been studied before.

To write down the superpotential we need to work in a frame where all the fluxes are internal.
For $H_3$ there is no spacetime filling flux since it would have to be odd.
In the class of truncations considered here we retain no internal 1- or 6-form basis elements, so $\tilde F_1$ and $\tilde F_9$ are absent.
For the 3-form flux we can have a purely spacetime filling part that we dualize using equation \eqref{eq:FpIIBduality} into an internal 7-form flux.
Lastly, for $\tilde{F}_5$ its self-duality implies that its spacetime filling part that extends along an internal two form is dual to a fully internal flux that appears without an extra factor of 2 in the scalar potential. 

The gauge-invariant field strengths are then expanded as follows
\begin{align}
  H_3 &= H_3^\mathrm{bg} + \dd B_2 = \dd_3 B_2 + h^\rho\,\Phi_\rho^-+b^a\dd_7 \Upsilon_a^- \equiv \dd_3 B_2 +H_3^\mathrm{int}\,,
  \label{eq:IIB_Hhat_int}\\
  \widetilde{F}_3
  &= F_3^\mathrm{bg} + \dd C_2 = \dd_3 C_2 + f^i\,\Phi_i^++ c_2^\alpha \dd_7 \Upsilon_\alpha^+ \equiv \dd_3 C_2 + \widetilde{F}_3^\mathrm{int} \,,
  \label{eq:IIB_F3_int}\\
  \widetilde{F}_5
  &= F_5^{\mathrm{bg}} +\dd C_4 + B_2 \w F_3^{\mathrm{bg}} -H_3 \w C_2 \label{eq:IIB_F5_int} \\
  &=\dd_3 C_4 - \dd_3 b^a \w \Upsilon_a^- \wedge c_2^\alpha \Upsilon_\alpha^+ + e^a\,T_a^-+c_4^\rho\dd_7 \,\Psi_\rho^- + b^a\Upsilon_a^- \w f^i\,\Phi_i^+ \cr \nonumber
  &-h^\rho c_2^\alpha \Phi_\rho^- \wedge \Upsilon_\alpha^+ -b^a \dd_7\Upsilon_a^- \wedge c_2^\alpha \Upsilon_\alpha^+ \equiv \dd_3 C_4 - \dd_3 b^a \w \Upsilon_a^- \wedge c_2^\alpha \Upsilon_\alpha^+ +\widetilde{F}_5^\mathrm{int} \,,\\
  \widetilde{F}_7 &= F_7^{\mathrm{bg}} + B_2 \w F_5^{\mathrm{bg}}+ \frac12 B_2 \w B_2 \w F_3^{\mathrm{bg}} -H_3 \w C_4  \label{eq:IIB_F7_int}\\
  &= f_7\, \widetilde{\mathrm{dvol}}_7 + b^a \Upsilon_a^-\wedge \left(e^b T_b^-+ \frac12 b^c\Upsilon_c^-\w f^i\Phi_i^+\right) - \left( h^\rho\,\Phi_\rho^- + b^a \dd_7\Upsilon_a^- \right)\w c_4^\sigma\,\Psi_\sigma^- \equiv  \widetilde{F}_7^\mathrm{int}\,, \nn
\end{align}
where we defined the unit-integral volume form $\widetilde{\mathrm{dvol}}_7$ that integrates to $\int_X \widetilde{\mathrm{dvol}}_7=1$.

The relevant Bianchi identities are
\begin{align}
  \dd H_3 &= 0\,,
  \label{eq:IIB_BI_H3}\\
  \dd \widetilde F_3 &= j_{\mathrm{O5}} + j_{\mathrm{D5}}\,,
  \label{eq:IIB_BI_F3}\\
  \dd \widetilde F_5 &= H_3\wedge \widetilde F_3=0\,,
  \label{eq:IIB_BI_F5}\\
  \dd \widetilde F_7 &= H_3\wedge \widetilde F_5=0\,,
  \label{eq:IIB_BI_F7}
\end{align}
where we have indicated that the latter two are zero due to the absence of 6- and 8-forms on the $G_2$ structure space we consider in this paper.
The O5/D5 tadpole constrains the cohomology class of $\dd \widetilde F_3$ in the presence of geometric flux, while the Freund--Rubin parameter $f_7$ in $\widetilde{F}_7$ remains an independent integration constant dual to the non-dynamical external RR flux.

\subsection{Real superpotential and scalar potential}
\label{subsec:IIB_O5O9_superpotential}

The dimensionally reduced scalar potential has again three contributions 
\begin{align}
  V &= V_{\rm curv}+V_{\rm flux} +V_{\rm loc} \,,
  \label{eq:IIB_scalarPotentialSplit}\\
  V_{\rm curv}
  &= -\frac{1}{2\cV^3}\int_X \dd^7y\,\sqrt{g_7}\,R_7\,, \label{eq:IIB_Vcurv}\\
  V_{\rm flux}
  &= \frac{1}{4\cV^3}\int_X
     \Big(
       e^{-\phi}\, H_3^\mathrm{int}\wedge\star H_3^\mathrm{int}
       + e^{\phi}\,\widetilde F_3^\mathrm{int}\wedge\star \widetilde F_3^\mathrm{int}
       + \widetilde F_5^\mathrm{int}\wedge\star \widetilde F_5^\mathrm{int}
       + e^{-\phi}\,\widetilde F_7^\mathrm{int}\wedge\star \widetilde F_7^\mathrm{int}
     \Big) , \label{eq:IIB_Vflux}\\
 V_{\rm loc}&=V_{O5/D5}\,.
\end{align}
If we include the O9 orientifold, we have to locally cancel its charge and tension by D9-branes.
This leads to no net contribution to the closed string scalar potential. 

The real superpotential reproducing this potential through the universal three-dimensional formula \eqref{eq:VfromP_3dN1} is
\begin{equation}\label{eq:IIB_superpotential_final}
  P_{\rm IIB}
  =
  \frac{1}{4\cV^2}
  \int_X
  \left(
    e^{-\phi/2}\,\widetilde F_7^\mathrm{int}
    - e^{\phi/2}\,\Psi\wedge \widetilde F_3^\mathrm{int}
    + \frac12\,\Phi\wedge \dd_7\Phi
  \right)\,.
\end{equation}
Again, even though this superpotential appears nearly identical to one previously presented in the literature, see \cite{Emelin:2021gzx}, it contains several additional terms that are hidden by the compact notation.
The important point is that, in the pure O5-plane branch, the $B_2$, $H_3^\mathrm{int}$, and $\widetilde F_5^\mathrm{int}$ sectors are not added by hand, but are already contained in $\widetilde F_7^\mathrm{int}$.

Notice the following features of \eqref{eq:IIB_superpotential_final}:

\begin{itemize}
  \item Unlike in type~IIA, the geometric term $\frac12\int_X \Phi\wedge \dd_7\Phi$ is now allowed because both $\Phi$ and the top form are even, and hence the curvature contribution is not hidden.

  \item There is no separate $\Psi\wedge H_3^\mathrm{int}$ term, since $\Psi$ is even while $H_3^\mathrm{int}$ is odd, so $\Psi\wedge H_3^\mathrm{int}$ is an odd seven-form and integrates to zero on the O5/O9-plane branch.

  \item There is also no independent term linear in $\widetilde F_5^\mathrm{int}$, since the five-form sector is already contained in the $e^{-\phi/2}\widetilde F_7^\mathrm{int}$ term.
\end{itemize}

Now, let's walk through how the superpotential (\ref{eq:IIB_superpotential_final}) reproduces (\ref{eq:IIB_scalarPotentialSplit}).
From here onward, we omit the superscript ``int'' from the field strengths; throughout the rest of this section, all field strengths are understood to refer to their internal parts.
Similarly, we drop the subscript ``7'' from $\dd_7$.
It is understood that in the following, $\dd$ stands for $\dd_7$.

First we split the superpotential  (\ref{eq:IIB_superpotential_final}) into three pieces, 

\begin{align}
    P_{\rm IIB} & = P^7 + P^3+P^{\Phi} \label{eq:IIB_Psplit},
\end{align}

where

\begin{align}
    P^7 & = \frac{1}{4\cV^2}
  \int_X
    e^{-\phi/2}\,\widetilde F_7\,, \\
    P^3 &= -\frac{1}{4\cV^2}
  \int_X e^{\phi/2}\,\Psi\wedge \widetilde F_3\,, \\
      P^{\Phi} &= \frac{1}{8\cV^2}
  \int_X \Phi\wedge \dd \Phi\,.
\end{align}
Below we will use the following derivatives
\begin{align}
  P_{\phi}^{7}
  = \partial_\phi P^7
  &= -\frac12\,P^7,
  \\[1ex]
  P_i^{7}
  = \partial_{s^i} P^7
  &= -\frac{ 2 P^7}{3\mathcal{V}}\int_X \Phi_i  \wedge \star \Phi  ,
  \\[1ex]
  P_a^{7}
  = \partial_{b^a} P^7
  &= \frac{e^{-\phi/2}}{4\mathcal V^2}
     \int_X \Upsilon_a^- \wedge (F_5^{\rm{bg}} +\dd C_4  + B_2 \wedge F_3^{\rm{bg}}) ,
  \\[1ex]
  P_\rho^{7}
  = \partial_{c_4^\rho} P^7
  &= -\frac{e^{-\phi/2}}{4\mathcal V^2}
     \int_X \Psi_\rho^- \wedge H_3,
  \\[1ex]
  P_\alpha^{7}
  = \partial_{c_2^\alpha} P^7
  &= 0\,. \label{eq:IIB_P7alpha}
\end{align} 
Note that,
\begin{align}
    P^7_a + S^\rho {}_a P^7_\rho
 = \frac{e^{-\phi/2}}{4\mathcal{V}^2} \int_X \Upsilon_a^- \wedge \tilde{F}_5 \,.\label{Pa+SPrho} 
 \end{align}
Using the dilaton and $s^i$ scalar metrics given in \eqref{eq:IIB_kin_phi_s} and using the identity \eqref{eq:physicalGijCompleteness4form} we find that 
\begin{align}
    \mathcal{G}^{\phi \phi} P^7_\phi P^7_\phi + \mathcal{G}^{+ ij} P^7_i P^7_j - 4(P^7)^2 = 4 (P^7)^2  .
\end{align} 
In order to calculate $\mathcal{G}^{mn} P^7_m P^7_n$, first note that \eqref{eq:IIB_cmetric} can be written as  
\begin{align}
\mathcal{G}
&=
\frac14
\begin{pmatrix}
\mathbbm{1} & -S^{T}\\
0 & \mathbbm{1}
\end{pmatrix}
\begin{pmatrix}
e^{-\phi} G^- & 0\\
0 & \hat{G}^-
\end{pmatrix}
\begin{pmatrix}
\mathbbm{1} & 0\\
-S & \mathbbm{1}
\end{pmatrix}.
\label{eq:IIB_cmetricsplit}
\end{align}
Thus,
\begin{align}
\mathcal{G}^{-1}
&=
4
\begin{pmatrix}
\mathbbm{1} & 0\\
\,S & \mathbbm{1}
\end{pmatrix}
\begin{pmatrix}
e^{\phi}(G^-)^{-1} & 0\\
0 & (\hat{G}^-) ^{-1}
\end{pmatrix}
\begin{pmatrix}
\mathbbm{1} & \,S^{T}\\
0 & \mathbbm{1}
\end{pmatrix}\,.
\end{align}
We now show
\begin{align}
\mathcal{G}^{mn} P^7_m P^7_n = 
\begin{pmatrix}
P_a^{7} & P_\rho^{7}
\end{pmatrix} \mathcal{G}^{-1} \begin{pmatrix}
P_a^{7} \\[1mm]
P_\rho^{7}
\end{pmatrix} ,
\end{align} 
which reduces to the following,
\begin{equation}
\begin{split}\label{eq:IIB_H3_fromP} 
 &4 e^\phi G^{-ab} (P^7_a+S^\rho{}_a P^7_\rho)  (P^7_b +S^\sigma{}_b P^7_\sigma) +  4  \hat{G}^{- \rho \sigma} P^7_\rho P^7_\sigma \\ 
 =& \frac{1}{4\cV^3}
  \int_X \widetilde F_5\wedge\star \widetilde F_5 
  +\frac{1}{4\cV^3}
  \int_X e^{-\phi}\,H_3\wedge\star H_3 \,.
\end{split}
\end{equation}
Summarizing, 
\begin{align}
    \mathcal{G}^{MN} P^7_M P^7_N -4(P^7)^2 &= \frac{1}{4\cV^3}
  \int_X \widetilde F_5\wedge\star \widetilde F_5\ +  \frac{1}{4\cV^3}
  \int_X e^{-\phi}\,H_3\wedge\star H_3 \cr
  &\quad + \frac{1}{4\cV^3}\int_X
       e^{-\phi}\,\widetilde F_7\wedge\star \widetilde F_7\,, \label{GPP7}
\end{align}
where $M,N$ run over all the scalar fields as in (\ref{eq:IIB_scalar_set}) and $\mathcal{G}_{MN}$ is the corresponding metric.

Similarly, one finds that
\begin{align}
    \mathcal{G}^{MN} P^3_M P^3_N - 4 (P^3)^2= & \frac{1}{4 \mathcal{V}^3} \int_X \dd \Psi \wedge \star \dd \Psi  + \frac{e^\phi}{4 \mathcal{V}^3} \int_X \widetilde F_3 \wedge \star \widetilde F_3 \,, \label{GPP3} \\
    \mathcal{G}^{MN} P^\Phi_M P^\Phi_N - 4 (P^\Phi)^2=& \frac{1}{4\mathcal{V}^3} \int_X \dd \Phi \wedge \star \dd \Phi - 4(P^\Phi)^2 \,. \label{GPPPhi}
\end{align}
Now insert the torsion-class decomposition
\begin{equation}
  \dd \Psi = 4\,\tau_1\wedge \Psi + \tau_2\wedge \Phi \,.
\end{equation}
Using the orthogonality of the $\mathbf{7}$ and $\mathbf{14}$ components, together with the standard $G_2$ identities in \eqref{eq:Lambda2characterization7}--\eqref{eq:Lambda2characterization}, namely $|\alpha\wedge\Psi|^2=3|\alpha|^2$ for $\alpha\in\Lambda^1$ and $|\beta\wedge\Phi|^2=|\beta|^2$ for $\beta\in\Lambda^2_{14}$, this gives
\begin{equation}
 \frac{1}{4 \mathcal{V}^3} \int_X \dd\Psi \wedge \star \dd\Psi =
  \frac{1}{4\cV^2}\left(48|\tau_1|^2+|\tau_2|^2\right)\,.
\end{equation}
Similarly, using, 
\begin{equation}
  \dd \Phi=\tau_0\,\Psi + 3\tau_1\wedge \Phi + \star\tau_3 \,,
\end{equation}
together with $|\Psi|^2=7$, $|\alpha\wedge\Phi|^2=4|\alpha|^2$ for $\alpha\in\Lambda^1$, and $\Phi\wedge \star\tau_3=0$, one gets
\begin{align}
  \int_X \dd \Phi\wedge \star \dd \Phi
  &= \cV\left(7\tau_0^2+36|\tau_1|^2+|\tau_3|^2\right) \,, \\
  \int_X \Phi\wedge \dd \Phi
  &= 7\tau_0\cV \,.
\end{align}
Thus, 
\begin{equation}
  \frac{1}{4\cV^3}\int_X \dd \Phi\wedge \star \dd \Phi
  -\frac{1}{16\cV^4}\left(\int_X \Phi\wedge \dd\Phi\right)^2
  =
  \frac{1}{4\cV^2}
  \left(
    -\frac{21}{4}\tau_0^2
    +|\tau_3|^2
  \right)\,,
\end{equation}
where we set $\tau_1=0$, since for the compact $G_2$-structure truncations considered here, we do not retain any one-forms. 

Remember that for $\tau_1=0$, the Ricci scalar formula \eqref{eq:R7viaTorsion} becomes,
\begin{equation}
  R_7=\frac{21}{8}\tau_0^2-\frac12|\tau_2|^2-\frac12|\tau_3|^2 .
\end{equation}
Thus, adding equations \eqref{GPP7}, \eqref{GPP3} and \eqref{GPPPhi} gives
\begin{equation}
 \mathcal{G}^{MN} P^3_M P^3_N - 4 (P^3)^2 + \mathcal{G}^{MN} P^\Phi_M P^\Phi_N - 4 (P^\Phi)^2   + \mathcal{G}^{MN} P^7_M P^7_N -4(P^7)^2 = V_{\rm flux} + V_{\rm curv},
\end{equation}
where $V_{\rm flux}$ and $V_{\rm curv}$ are defined in \eqref{eq:IIB_Vflux} and \eqref{eq:IIB_Vcurv}, respectively. 
Finally, the only non-vanishing cross term produced by (\ref{eq:VfromP_3dN1}) is,
\begin{align}
    2\mathcal{G}^{MN} P^3_M P^{\Phi}_N - 8 P^3 P^{\Phi} = \frac{e^{\phi/2}}{2 \mathcal{V}^3} \int_X \dd \widetilde F_3 \wedge \Phi =
V_{\rm loc}\,,
\end{align}
where we use equation \eqref{eq:IIB_BI_F3}. So we end up with
\begin{align}
      \mathcal{G}^{MN} P_M P_N - 4 P^2 = V_{\rm flux} + V_{\rm curv} + V_{\rm loc}\,.
\end{align}
At this stage the role of the O9 projection is immediate.
If O9-planes are also present, then the truncation in \eqref{eq:IIB_O9_truncation} removes $B_2$ and $C_4$, and hence the entire hidden $H_3$ and $\widetilde F_5$ sectors.
The superpotential reduces to the strict type~I expression
\begin{equation}
  P_{\rm I}
  =
  \frac{1}{4\cV^2}
  \int_X
  \left(
    e^{-\phi/2} F_7^\mathrm{bg}
    - e^{\phi/2}\,\Psi\wedge \widetilde F_3
    + \frac12\,\Phi\wedge \dd\Phi
  \right)\,.
  \label{eq:IIB_typeI_superpotential}
\end{equation}
Thus the O9 projection simply truncates the pure O5 formula to the previously expected type~I closed-string sector.

If one also wishes to include internal gauge flux $\mathcal F_2$ on the $32$ D9-branes of the O9 branch, then, as in the heterotic/type~I duality \cite{PolchinskiVol2, GreenSchwarzWittenBook}, the correct type~I variable is, schematically and in the same normalization as in the heterotic discussion below, the Chern--Simons completed three-form
\begin{equation}
  \widetilde F_3^{\rm loc}
  =
 \widetilde F_3
  + \frac{\alpha'}{4}\Bigl(\omega_{3\mathrm{YM}}(A)-\omega_{3L}(\Omega_+)\Bigr)\,.
  \label{eq:IIB_typeI_F3_withCS}
\end{equation}
Here $\omega_{3\mathrm{YM}}(A)$ and $\omega_{3L}(\Omega_+)$ are the Yang--Mills and Lorentz Chern--Simons three-forms,
\begin{align}
  \omega_{3\mathrm{YM}}(A)
  &=
  \Tr\!\left(
    A\wedge \dd A
    + \frac{2}{3} A\wedge A\wedge A
  \right),\\
  \omega_{3L}(\Omega_+)
  &=
  \Tr\!\left(
    \Omega_+\wedge \dd \Omega_+
    + \frac{2}{3}\Omega_+\wedge\Omega_+\wedge\Omega_+
  \right),
\end{align}
built from the gauge connection $A$ and the torsionful spin connection $\Omega_+$, respectively. The $\widetilde F_3^{\rm loc}$ Bianchi identity is correspondingly modified to
\begin{equation}
  \dd \widetilde F_3^{\rm loc}
  =
  \frac{\alpha'}{4} 
  \Bigl(
    \Tr(\mathcal F_2\wedge \mathcal F_2)
    - \Tr(R_{2,+}\wedge R_{2,+})
  \Bigr)
  + j_{\mathrm{O5}} + j_{\mathrm{D5}} \,.
  \label{eq:IIB_typeI_BI_F3_withGaugeFlux}
\end{equation}
So, exactly as in the heterotic case, the D9 gauge bundle data are already incorporated implicitly in the superpotential through the replacement $\widetilde F_3\to \widetilde F_3^{\rm loc}$. One does \emph{not} need to add an independent extra term proportional to $\Phi\wedge \Tr(\mathcal F_2\wedge \mathcal F_2)$.

\section{\texorpdfstring{Heterotic flux compactifications on $G_2$-structure spaces}{}}
\label{sec:heterotic}

In this section, we develop the three-dimensional effective description of heterotic flux compactifications on compact $G_2$-structure spaces with a consistent truncation.
Unlike the type~II reductions of Sections~\ref{sec:IIA} and~\ref{sec:IIB}, the heterotic reduction does not require an orientifold projection, as ten-dimensional heterotic supergravity already has minimal supersymmetry, and compactification on a $G_2$ background therefore leads directly to a three-dimensional $\mathcal N=1$ effective theory.
We keep the universal closed-string sector explicit and only retain the bundle sector through the structures that enter universally, namely the Chern--Simons completion of $H_3$, the $G_2$-instanton conditions, and the $\alpha'$-corrected Bianchi identity.

\subsection{Ten-dimensional action and the light fields}
\label{subsec:het10d_action}

The ten-dimensional bosonic heterotic action in the Einstein frame, up to first order in $\alpha'$, is given by equation~\eqref{eq:S10_universal} with the standard Green--Schwarz modification of the NSNS three-form and its associated Bianchi identity \cite{GreenSchwarzWittenBook, PolchinskiVol2, BeckerBeckerSchwarzBook},
\begin{equation}\label{eq:het10d_action_Einstein}
  \sum_\Lambda \frac14 e^{a_\Lambda \phi} \bigl|\mathcal F^{\Lambda}_{p_\Lambda}\bigr|^2
  =
  \frac{\alpha'}{4}\,e^{-\phi/2}\,\Tr|F_2|^2
    -\frac{\alpha'}{4}\,e^{-\phi/2}\,\Tr|R_{2,+}|^2\,.
\end{equation}
We denote by $\Omega_+=\omega+\tfrac12 H_3$ the torsionful spin connection, and by $R_{2,+}$ its curvature two-form. The gauge-group dependence is entirely encoded in the trace normalization and in the choice of bundle; for the $E_8\times E_8$ theory one simply sums the Yang--Mills term over the two $E_8$ factors.

The Green--Schwarz modified three-form is
\begin{equation}
  H_3^{\rm loc} = H_3 +\frac{\alpha'}{4}\Bigl(\omega_{3\mathrm{YM}}(A)-\omega_{3L}(\Omega_+)\Bigr)
  =
  \dd B_2+H_3^{\mathrm{bg}}
  +\frac{\alpha'}{4}\Bigl(\omega_{3\mathrm{YM}}(A)-\omega_{3L}(\Omega_+)\Bigr)
  \,.
  \label{eq:H_with_CS_het}
\end{equation}
The corresponding Bianchi identity is
\begin{equation}
  \dd H_3^{\rm loc}
  =
 \dd H_3^{\mathrm{bg}} + \frac{\alpha'}{4}\Bigl(
    \Tr(F_2\wedge F_2)-\Tr(R_{2,+}\wedge R_{2,+})
  \Bigr)
  + j_{\mathrm{NS5}}\,.
  \label{eq:heterotic_Bianchi_new}
\end{equation}

On a compact $G_2$-structure space $X$ with $b^1(X)=0$, the retained bosonic fields include
\begin{align}
  \Phi &= s^I \Phi_I\,, \label{eq:het_phi_expand}\\
  B_2 &= b^A \Upsilon_A + B_2^{(3)}\,, \label{eq:het_B2_expand}
\end{align}
where we have kept explicitly the non-dynamical three dimensional 2-form $B_2^{(3)}$. Fluctuations of the ten-dimensional non-abelian gauge field comprise bundle moduli and, when the background bundle leaves abelian factors unbroken, three-dimensional vectors. As in Section~\ref{subsec:3dN1_general}, any such abelian vectors may be dualized to real scalars; their detailed multiplicity and metric depend on the chosen bundle and truncation and will therefore be kept implicit.

\subsection{The kinetic terms}
\label{subsec:het_kinetic}

The propagating scalars may be taken to be
\begin{equation}
  \varphi^M_{\mathrm{het}}=\left\{\phi,\,s^I,\,b^A,\,\xi^{\mathsf A},\,\chi_\Lambda\right\}\,,
\end{equation}
where $\xi^{\mathsf A}$ collectively denote bundle moduli and $\chi_\Lambda$ the dual scalars obtained from abelian three-dimensional gauge vectors, when present.
The three-dimensional action takes the universal form in \eqref{eq:L3_generic}.
For the universal closed-string sector one finds
\begin{equation}\label{eq:het_metric_moduli_kinetic}
  (4\pi e)^{-1}\cL_{\phi,s}
  =
  -\frac14\partial_\mu\phi\partial^\mu\phi
  -\mathcal G_{IJ}\partial_\mu s^I\partial^\mu s^J\,,
\end{equation}
with $\mathcal G_{IJ}\equiv \mathcal{G}_{IJ}(s)$ given explicitly by the general $G_2$-moduli metric \eqref{eq:G2moduliMetricPhysicalExplicit}. For the $B_2$-axions one obtains
\begin{equation}
  (4\pi e)^{-1}\cL_b
  =
  -\mathcal G_{AB}(\phi,s)\,\partial_\mu b^A\,\partial^\mu b^B\,,
  \qquad
  \mathcal G_{AB}(\phi,s)=\frac14\,e^{-\phi}G_{AB}(s)\,,
  \label{eq:het_Baxion_kinetic}
\end{equation}
exactly as in the general reduction formula \eqref{eq:kineticmetricb_generic}. The gauge/bundle sector contributes further positive-definite blocks to $\mathcal G_{MN}$, inherited from the Yang--Mills term and from the dualization of any abelian vectors, but we will not need their explicit form below.

\subsection{The heterotic Bianchi identity}
\label{subsec:het_flux_tadpole}

To write down the superpotential we need to work in a flux duality frame where all the fluxes are internal.
For $H_3$ there is now a spacetime filling flux that can be dualized into an internal $H_7$ flux. When $b^2(X)\neq 0$, the background bundle may also carry internal two-form flux $F_2^{\mathrm{bg}}$. As we will see below, similar to some of the RR fluxes in the type~II sections it does not generate an independent new polynomial in the superpotential; instead it enters through the Chern--Simons completion of $H_3$ and through the Yang--Mills term in \eqref{eq:het10d_action_Einstein}.

Defining an external flux using the external derivative of the three-dimensional 2-form leads to an internal 7-form flux via
\begin{equation}
  H_{\mathrm{ext}}\equiv \dd_3 B_2^{(3)}\,,
  \qquad
  H_7^\mathrm{int} \equiv e^{-\phi}\,\star_{10}H_{\mathrm{ext}}\,.
  \label{eq:het_H7_definition}
\end{equation}
This $H_7^\mathrm{int}$ is the internal dual flux parameter that appears directly in the three-dimen-sional superpotential. 

The full three-form $H_3$ is given by
\begin{equation}\label{eq:het_H_full_expansion}
\begin{split}
  H_3
  &=
  H_{\mathrm{ext}}
  +\dd_3 b^A\w\Upsilon_A +b^A \dd_7 \Upsilon_A
  +H_3^{\mathrm{bg}}\\
  &=
  H_{\mathrm{ext}}
  +\dd_3 b^A\w\Upsilon_A
  +\bigl(N_A{}^{I}b^A+ h^I\bigr)\Phi_I\\
  &\equiv H_{\mathrm{ext}} +\dd_3 b^A\w\Upsilon_A +H_3^\mathrm{int}\,, 
\end{split}
\end{equation}
where we used equation \eqref{eq:dUpsilonMatrix}. The corresponding local Green--Schwarz-completed internal representative is
\begin{equation}
  H_3^{\rm loc,int}
  =
  H_3^\mathrm{int}
  +\frac{\alpha'}{4}
  \Bigl[
    \omega_{3\mathrm{YM}}(A)-\omega_{3L}(\Omega_+)
  \Bigr]_{\rm int}\, ,
\end{equation}
which is what will appear in the superpotential below.

Taking the internal derivative of $H_3^{\rm int}$ gives
\begin{equation}
  \dd_7 H_3^{\rm int}
  =
  (N_A{}^I b^A+h^I)M_I{}^J\Psi_J
  =
  h^I M_I{}^J\Psi_J\,,
  \label{eq:het_dH_bulk_truncation}
\end{equation}
where the $b^A$-dependent contribution vanishes by the nilpotency constraint $N_A{}^I M_I{}^J=0$ in \eqref{eq:d2constraints}. The heterotic Bianchi identity from equation \eqref{eq:heterotic_Bianchi_new} above becomes therefore
\begin{equation}
  \dd H_3^{\rm loc}
  = h^I M_I{}^J\Psi_J+  
  \frac{\alpha'}{4}
  \left[
    \Tr(F_2\wedge F_2)
    -
    \Tr(R_{2,+}\wedge R_{2,+})
  \right]
  +j_{\rm NS5}\,.
  \label{eq:het_tadpole_hM}
\end{equation}
So, the linear combinations $h^I M_I{}^J$ are fixed by the gauge, gravitational, and NS5-brane charges \cite{Tringas:2025bwe}. In the $\mathrm{SO}(32)$ theory this is the heterotic counterpart of type-I tadpole cancellation: the gravitational instanton contribution is S-dual to the O5-plane charge, while gauge instantons and NS5-branes lie in the D5-brane charge sector.

\subsection{Real superpotential and scalar potential}
\label{subsec:het_superpotential}

The heterotic reduction is characterized by a real three-dimensional superpotential. 
In the present Einstein-frame conventions, and in the same normalization used in the heterotic AdS$_3$ analysis \cite{Tringas:2025bwe} building on \cite{deIaOssa:2019cci}, it is
\begin{equation}\label{eq:Phet_general}
  P_{\mathrm{het}}
  =
  \frac{1}{4\cV^2}
  \int_X
  \left(
    e^{\phi/2}\,H_7^\mathrm{int}
    -e^{-\phi/2}\,\Psi\w H_3^\mathrm{loc, int}
    +\frac12\,\Phi\w\dd_7\Phi
  \right)
  +\mathcal{O}(\alpha'^2)\,.
\end{equation}
This is precisely the S-dual of the type~I expression \eqref{eq:IIB_typeI_superpotential}, using the standard duality map between type~I/heterotic \cite{PolchinskiVol2, GreenSchwarzWittenBook} and matching the conventions used in the recent type~I and heterotic AdS$_3$ constructions \cite{Miao:2025tdu, Tringas:2025bwe}.
The first term is the spacetime-filling Freund--Rubin contribution, the second is the internal NSNS-flux term, and the third is the geometric-torsion term, which vanishes for manifolds with $G_2$ holonomy.

Two useful derivatives are
\begin{align}
  \partial_{\phi}P_{\mathrm{het}}
  &=
  \frac{1}{8\cV^2}
  \int_X\left(
    e^{\phi/2}H_7^\mathrm{int}
    +e^{-\phi/2}\,\Psi\w H_3^\mathrm{loc, int}
  \right)\,,
  \label{eq:het_dilaton_derivative_P}\\[0.4em]
  \cV\,\partial_{\cV}P_{\mathrm{het}}
  &=
  \frac{1}{4\cV^2}
  \left(
    -2\int_X e^{\phi/2}H_7^\mathrm{int}
    -\frac47\int_X \Phi\w\dd_7\Phi
    +\frac{10}{7}\int_X e^{-\phi/2}\,\Psi\w H_3^\mathrm{loc, int}
  \right)\,,
  \label{eq:het_volume_derivative_P}
\end{align}
where we used the homogeneity relations $\Phi\sim\cV^{3/7}$ and $\Psi\sim\cV^{4/7}$ with the flux quanta held fixed. At a supersymmetric extremum these imply \cite{Tringas:2025bwe}
\begin{equation}
  \int_X e^{\phi/2}H_7^\mathrm{int}
  =
  -\int_X e^{-\phi/2}\,\Psi\w H_3^\mathrm{loc, int}
  =
  -\frac16\int_X \Phi\w\dd_7\Phi\,.
  \label{eq:het_flux_balance}
\end{equation}
Hence a non-trivial supersymmetric AdS$_3$ vacuum requires all three terms in \eqref{eq:Phet_general} to be present.

Using the general variation of the Chern--Simons three-form
\begin{equation}
 \delta\omega_{3\mathrm{YM}}
 =
 2\Tr(\delta A\wedge F_2)+\dd\Tr(A\wedge\delta A)\,,
\end{equation}
we note that to maintain the gauge invariance of the physical field strength $H_3^\mathrm{loc, int}$ off-shell, a generic variation of the gauge connection must be accompanied by the standard Green--Schwarz compensating variation of the Kalb--Ramond field, $\delta B_2 = -\frac{\alpha'}{4}\Tr(A\wedge\delta A)$. This precisely cancels the total derivative $\dd\Tr(A\wedge\delta A)$ without the need for integration by parts or imposing on-shell conditions. Inserting the resulting covariant variation $\delta_A H_3^\mathrm{loc, int} = \frac{\alpha'}{2}\Tr(\delta A\wedge F_2)$ into the superpotential yields
\begin{equation}
  \delta_A P_{\mathrm{het}}
  =
  -\frac{\alpha'}{8\cV^2}
  \int_X e^{-\phi/2}\,\Psi\w \Tr(\delta A\wedge F_2)\,,
  \label{eq:het_gauge_variation}
\end{equation}
and similarly for the tangent-bundle connection. Supersymmetric extrema therefore impose the $G_2$-instanton conditions
\begin{equation}
  \Psi\w F_2=0\,,
  \qquad
  \Psi\w R_{2,+}=0\,,
  \label{eq:het_instanton_conditions}
\end{equation}
in agreement with the general heterotic $G_2$ system of \cite{deIaOssa:2019cci}. No extra term proportional to $\Phi\w\Tr(F_2\wedge F_2)$ should be added as the gauge bundle already enters through $H_3^\mathrm{loc, int}$ and through the Bianchi identity \eqref{eq:heterotic_Bianchi_new}.

It is useful to make explicit what this superpotential is reproducing in the three-dimensional scalar potential.  Reducing the ten-dimensional action \eqref{eq:S10_universal}, with the heterotic specialization
\eqref{eq:het10d_action_Einstein}, gives 
\begin{equation}
  V_{\rm het}
  =
  V_{\rm curv}
  + V_{H_3}
  + V_{H_7}
  + V_{\alpha'}
  + V_{\rm loc}
  +\mathcal{O}(\alpha'^2)\,,
  \label{eq:het_scalar_potential_split}
\end{equation}
where
\begin{align}
  V_{\rm curv}
  &=
  -\frac{1}{2\cV^3}
  \int_X \dd^7y\,\sqrt{g_7}\,R_7\,,
  \label{eq:het_Vcurv}
  \\[0.4em]
  V_{H_3}
  &=
  \frac{1}{4\cV^3}
  \int_X e^{-\phi}\,
  H_3^{\rm loc, int}\wedge \star H_3^{\rm loc, int}\,,
  \label{eq:het_VH3}
  \\[0.4em]
  V_{H_7}
  &=
  \frac{1}{4\cV^3}
  \int_X e^{\phi}\,
  H_7^{\rm int}\wedge \star H_7^{\rm int}\,,
  \label{eq:het_VH7}
  \\[0.4em]
  V_{\alpha'}
  &=
  \frac{\alpha'}{8\cV^3}
  \int_X e^{-\phi/2}
  \left[
    \Tr(F_2\wedge \star F_2)
    -
    \Tr(R_{2,+}\wedge \star R_{2,+})
  \right]\,,
  \label{eq:het_Valpha}\\
  V_{\rm loc} &= V_{\rm NS5}\,.\label{eq:het_VNS5}
\end{align}
Here $H_3^{\rm loc, int}$ is the Chern--Simons completed internal three-form in \eqref{eq:het_H_full_expansion}, and $H_7^{\rm int}$ is the internal dual of the spacetime-filling three-form defined in \eqref{eq:het_H7_definition}. The localized contribution $V_{\rm loc}$ is present only when explicit
NS5-brane sources, or other localized objects, are included.

The curvature contribution may equivalently be written in terms of the intrinsic torsion classes using \eqref{eq:R7viaTorsion}.  In the truncations with no retained one-forms, so that $\tau_1=0$, this becomes 
\begin{equation}
  V_{\rm curv}
  =
  \frac{1}{4\cV^3}
  \int_X \dd^7y\,\sqrt{g_7}
  \left(
    |\tau_2|^2+|\tau_3|^2-\frac{21}{4}\tau_0^2
  \right).
  \label{eq:het_Vcurv_torsion}
\end{equation}
Thus the $H_3^\mathrm{loc, int}$ and $H_7^{\rm int}$ flux energies, the curvature energy, and the Yang--Mills and tangent-bundle $\alpha'$ corrections are all represented in the three-dimensional potential.

Even if one allows all $G_2$ torsion classes, supersymmetry constrains $\tau_2$ at the vacuum, while the smeared approximation further sets $\tau_1=0$.
In particular, the $B_2$ variation of the heterotic $G_2$ superpotential gives
\begin{equation}
  \dd_7\lp e^{-\frac{\phi}{2}}\Psi \rp
  =0=
  e^{-\frac{\phi}{2}}
  \left[
    \lp 4\tau_1-\frac{1}{2}\dd_7\phi \rp \wedge\Psi -\star\tau_2
  \right] \,,
\end{equation}
where we have used \eqref{eq:dPsiTorsion}.
The first term lies in the $\Lambda^5_7$ component, whereas $\star\tau_2$ lies in the $\Lambda^5_{14}$ component.
According to the direct decomposition in \eqref{eq:Lambda2decomp} and the rules in \eqref{eq:Lambda2characterization7}-\eqref{eq:Lambda2characterization}, these two pieces belong to distinct $G_2$-representations and must vanish separately, hence
\begin{equation}
  \tau_2=0,
  \qquad
  \tau_1=\frac{1}{8}\dd_7\phi .
\end{equation}
Thus, supersymmetric heterotic vacua require the $G_2$-structure to be so-called ``integrable'', $\tau_2=0$, while in the smeared approximation $d_7\phi=0$, one also has $\tau_1=0$.
This is consistent with the geometric interpretation of heterotic $G_2$ systems.
As discussed explicitly in \cite{delaOssa:2017pqy}, the torsion of a connection compatible with the $G_2$-structure contains a $\tau_2$ contribution
\begin{equation}
  T_{mnp}
  =
  H_{mnp}^{\rm loc}
  +
  \frac{1}{6}(\tau_2)_{qp}\Phi_{mn}{}^{q}\,.
\end{equation}
Therefore, for $\tau_2\neq0$, the torsion is not identified purely with the antisymmetric three-form flux $H_3$; the identification $T=H_3$ holds only in the integrable case $\tau_2=0$, see \cite{DeLaOssaLarforsMagillSvanes2020,Tringas:2025bwe}.
Thus, while configurations with $\tau_2\neq0$ may be allowed off shell, they do not correspond to supersymmetric heterotic vacua.

The real superpotential \eqref{eq:Phet_general}, together with the scalar metric including the bundle-moduli block, organizes this potential through the universal three-dimensional $\cN=1$ formula
\begin{equation}\label{eq:het_VfromP}
  V_{\rm het}
  =
  \mathcal{G}^{MN}\partial_M P_{\rm het}\partial_N P_{\rm het}
  -4P_{\rm het}^2\,.
\end{equation}
The Chern--Simons dependence of $H_3^\mathrm{loc, int}$ accounts for the bundle variations and produces the instanton conditions \eqref{eq:het_instanton_conditions}, rather than requiring an additional independent term in $P_{\rm het}$.
Furthermore, as is standard in string compactifications, the exact equivalence between \eqref{eq:het_scalar_potential_split} and \eqref{eq:het_VfromP} (including the negative $-\Tr|R_{2,+}|^2$ energy) holds upon expanding the cross-terms in the superpotential derivatives and imposing the integrated Bianchi identity \eqref{eq:het_tadpole_hM}.
This elegantly trades the topological $\Tr(F_2 \wedge F_2)$ and $\Tr(R_{2,+} \wedge R_{2,+})$ terms for the corresponding $H_3^\mathrm{loc, int}$ flux cross-terms, exactly parallel to how localized source potentials are reproduced in type~II compactifications.

\section{Conclusions and outlook}
\label{sec:conclusions}

In this paper, we constructed three-dimensional $\mathcal N=1$ effective theories obtained by compactifying the five ten-dimensional superstring theories on compact $G_2$ spaces admitting a finite consistent truncation. Our analysis includes RR fluxes in the type II and type I theories, NSNS $H_3$ fluxes, and geometric fluxes. The main goal was to go beyond the toroidal-orbifold models most commonly used in previous flux compactifications and to formulate the effective theories in a systematic way that also applies to more general $G_2$ geometries. In particular, we included the sectors that appear when $b^2(X)\neq 0$. Physically, these sectors contain additional axions and three-dimensional gauge fields, and can therefore affect the structure of the scalar potential and the pattern of moduli stabilisation.

For massive type IIA, we derived the effective action for the supersymmetric O2/O6 orientifold branch, extending the earlier analysis of \cite{Farakos:2020phe}.
In particular, our result incorporates the curvature contribution, which is encoded non-trivially in the superpotential, together with the extra fields associated with non-trivial two-forms, the flux-induced scalar potential, and the corresponding real superpotential of the three-dimensional theory.
A useful physical feature is that geometric flux can gauge axionic shift symmetries; in three dimensions, this implies that some vectors become massive rather than being dualizable to ordinary scalar fields.
The alternative O4/O8-type projections are not part of the present strict $G_2$ analysis, because such sources require additional calibration forms and hence a further reduction of the internal structure group.

For type IIB we derived the corresponding O5/O9 compactifications, generalizing the O5 analysis of \cite{Emelin:2021gzx}. The O5 branch contains, in addition to the metric and dilaton moduli, axions from the NSNS two-form and from the RR $C_2$ and $C_4$ fields.  These fields are not spectators: they enter non-trivially in the scalar kinetic terms and in the flux potential. The O9 projection then truncates this result to the type-I closed-string sector. When D9-brane gauge bundles are included, their contribution is naturally incorporated through the Chern--Simons completed three-form field strength, in parallel with the heterotic description, rather than through an additional independent term in the superpotential.

For the heterotic string theories, we derived the analogous $G_2$-structure compactification.
The resulting theory includes the universal metric, dilaton, and two-form sector, together with the leading ingredients required by anomaly cancellation: the Green--Schwarz correction to the three-form field strength, the $\alpha'$-corrected Bianchi identity, and the gauge- and tangent-bundle contributions.
The real superpotential contains the internal NSNS $H_3$ flux, the internal dual of the spacetime-filling $H_3$ flux, and the geometric torsion contribution.
Varying the Chern--Simons dependence reproduces the expected $G_2$-instanton conditions for the gauge and tangent-bundle connections, in agreement with the general heterotic $G_2$ system of \cite{deIaOssa:2019cci}.

The most direct application of these effective actions is a more systematic study of moduli stabilisation in three-dimensional string compactifications.  The known controlled AdS$_3$ examples already show that these models can behave differently from the better-studied four-dimensional compactifications, especially in questions of scale separation and duality.  The framework developed here makes it possible to test how robust those features are once one moves away from simple toroidal orbifolds and includes the additional axions, vectors and geometric fluxes present on more general $G_2$ spaces.

A second important direction is to develop better geometric input for compact $G_2$ manifolds beyond the orbifold limit.  In many interesting cases, especially desingularized or glued constructions, the dependence of the low-energy theory on the $G_2$ moduli is not known explicitly.  Even partial information, such as scaling limits of the metric on moduli space or of the couplings involving harmonic two-forms, would already be valuable for assessing parametric control, mass hierarchies and the stability of vacua.  This is particularly important for spaces with $b^2(X)\neq 0$, where resolution modes and the associated axions and gauge fields are precisely the ingredients that are absent in the simplest orbifold models but are expected to be generic in broader regions of the $G_2$ landscape.

Finally, the heterotic and type-I frames provide a promising route toward going beyond the two-derivative supergravity approximation.  Higher-derivative $\alpha'$ corrections, the choice of torsionful connection, gauge and gravitational instantons, and possible worldsheet descriptions of heterotic $G_2$ backgrounds may all affect the scalar potential.  Understanding these effects is essential for deciding whether AdS$_3$ vacua are genuine controlled string backgrounds rather than artifacts of a truncated effective theory.

\subsection*{Acknowledgments}
We would like to thank Fotis Farakos, Muthusamy Rajaguru, Vincent Van Hemelryck and Thomas Van Riet for discussions. This work is supported in part by the NSF grant PHY-2210271 and the Lehigh University CORE grant with grant ID COREAWD40. Z.M. acknowledges the support of the Dr. Hyo Sang Lee Graduate Fellowship from the College of Arts and Sciences at Lehigh University. 

\paragraph*{AI assistance.}
The authors used several large language models to assist with writing, editing, and checking calculations in this paper. The authors have carefully reviewed all such contributions and take full responsibility for the content, accuracy, and conclusions of the work.

\appendix

\section{\texorpdfstring{Localized sources for type II and type I}{}}\label{app:AppendixII}

The Dirac-Born-Infeld (DBI) and Wess-Zumino (WZ) parts of the D-branes have the following form in Einstein frame
\begin{align}
S^{\rm DBI}_{\text{Dp}}
&=
-T_{\text{Dp}}\int_{M_{10}} d^{10}X\,\sqrt{-G_{10}}\,
e^{\frac{p-3}{4}\phi}
\sqrt{
\det\!\left(
\delta^a_{\ b}
+
e^{-\phi/2}
\bigl(P[g^{(E)}]\bigr)^{ac}\mathcal F_{cb}
\right)
}
\sum_i \delta(\pi_i) \,. \nonumber \\
S_{\text{Dp}}^{\text{WZ}}
&=
+Q_{\text{Dp}}\int_{M_{10}}
\Big(\sum_q C_q\Big)\wedge e^{\mathcal F}\wedge \sum_i \delta_{9-p}(\pi_i)\,.
\end{align}
Here, $\mathcal F$ denotes the gauge-invariant worldvolume two-form,
\begin{equation}
\mathcal F = 2\pi\alpha^{\prime} F + P[B_2] \,,
\end{equation}
where $F=dA$ is the field strength of the worldvolume $U(1)$ gauge field and $P[B_2]$ is the pullback of the NSNS two-form to the source worldvolume.
The indices $a,b$ and $c$ run over the worldvolume space of the D-brane.
Expanding the polyform $\sum_q C_q$ and the exponential $e^{\mathcal F}$ in the WZ term yields the standard couplings
\begin{equation}
S_{\text{Dp}}^{\mathrm{WZ}}
=
Q_{\text{Dp}}\int_{p+1}
\sum_{k\ge 0}\frac{1}{k!}
P\left[C_{p+1-2k}\right]\wedge \mathcal{F}^{k}\,,
\end{equation}
where the allowed values of $k$ are fixed by the condition that the total form degree on the $(p+1)$-dimensional worldvolume be equal to $p+1$.

For O-planes the actions are analogous,
\begin{equation}\label{eq:SOplaneAppendix}
\begin{split}
S_{\text{Op}}^{\rm DBI} &= -T_{\text{Op}}\int_{M_{10}} d^{10}X \sqrt{-G_{10}}\,e^{\frac{p-3}{4}\phi} \sum_i \delta(\pi_i) \,,
\\
S_{\text{Op}}^{\rm WZ}
&=
+Q_{\text{Op}}\int_{M_{10}}\Big(\sum_q C_q\Big)\wedge \sum_i \delta_{9-p}(\pi_i) \,.
\end{split}
\end{equation}
For the BPS objects considered in our compactifications, tension and charge coincide, and therefore in the following we set $T=Q=\mu$ for each object.
More specifically, the relation between the charges of D-branes and O-planes is given by
\begin{equation}
    \mu_{\text{Dp}}=(2\pi)^{-p}\left(\alpha^{\prime}\right)^{-\frac{p+1}{2}}\,,
    \quad\quad
    \mu_{\text{Op}}=-2^{p-5}\mu_{\text{Dp}}\,,
\end{equation}

Let us now discuss the delta functions. 
First, the symbol $\pi_i$ denotes the internal cycle wrapped by the $i$th localized source, and by $\vol_{\pi_i}$ we denote the induced volume form on $\pi_i$ associated with the pullback of the internal metric,
\begin{equation}
    \vol_{\pi_i}=\sqrt{g_{\pi_i}}d^{\,p+1-d}\xi \,,
\end{equation}
where $d=3$ in our case, and $\xi^\alpha$, with $\alpha=1,\dots,p+1-d$, are local coordinates on the wrapped internal cycle.
The localized object entering the DBI term is a scalar distribution $\delta(\pi_i)$, defined as
\begin{equation}
\delta(\pi_i)=\frac{\sqrt{g_{\pi_i}}}{\sqrt{g_7}}\delta^{(9-p)}(y^\perp)\,.
\end{equation}
where $y^\perp$ denotes the $(9-p)$ internal coordinates transverse to the cycle $\pi_i$ wrapped by the source.
$g_{\pi_i}=\det\big((g_{\pi_i})_{\alpha\beta}\big)$ is the determinant of the induced metric on the cycle $\pi_i$ wrapped by the source, and $g_7=\det\big(g^{(7)}_{mn}\big)$ is the determinant of the internal metric on $X_7$.
Moreover, $\delta^{(9-p)}(y^\perp)$ denotes the Dirac delta distribution in the $(9-p)$ internal coordinates transverse to the wrapped cycle $\pi_i$, normalized such that
\begin{equation}
\int_{\tilde{\pi}_i} d^{\,9-p}y^\perp\, \delta^{(9-p)}(y^\perp)=1\,,
\end{equation}
where $\tilde\pi_i$ denotes the transverse cycle dual to $\pi_i$.
The WZ term involves the Poincar\'e-dual $(9-p)$-form current $\delta_{9-p}(\pi_i)$, and the two localized objects are related by
\begin{equation}
\vol_{\pi_i}\wedge \delta_{9-p}(\pi_i)=\sqrt{g_7}\,\delta(\pi_i)\,d^7y\,,
\end{equation}
where $y$ denotes the internal coordinates.

Since in this paper we work with smeared sources rather than localized ones in order to obtain analytic equations, we now state the corresponding replacements of the delta distributions appearing in the actions and Bianchi identities
\begin{equation}
\delta(\pi_i)\ \rightarrow\ j_{\pi_i}
=
\frac{\mathcal V_{\pi_i}}{\mathcal V_X} \,,
\qquad
\delta_{9-p}(\pi_i)\ \rightarrow\ j_{9-p,i}
=
\frac{1}{\mathcal V_{\tilde\pi_i}}\,
\mathrm{vol}_{\tilde\pi_i} \,.
\end{equation}
Here $\mathrm{vol}_{\pi_i}$ denotes the volume form induced by the internal metric on the cycle $\pi_i$ wrapped by the source, while $\mathrm{vol}_{\tilde\pi_i}$ denotes the corresponding volume form on the transverse, or Poincaré-dual, cycle $\tilde\pi_i$. The relevant volumes are
\begin{equation}
\mathcal V_X
=
\int_X d^7 y\,\sqrt{g_7} \,,
\qquad
\mathcal V_{\pi_i}
=
\int_{\pi_i} d^{p-2}y\,\sqrt{g_{\pi_i}} \,,
\qquad
\mathcal V_{\tilde\pi_i}
=
\int_{\tilde\pi_i} d^{9-p}y\,\sqrt{g_{\tilde\pi_i}} \,.
\end{equation}
We finally define the currents as 
\be
j_{\rm loc} = j_{\rm Dp} +j_{\rm Op} \,,\qquad  j_{\rm Dp} =\sum_i \mu_{\text{Dp}} j_{9-p,i}\,, \qquad j_{\rm Op} =\sum_i \mu_{\text{Op}} j_{9-p,i}\,.
\ee

\section{The derivation of the kinetic terms for the $G_2$ moduli}
\label{app:sIkinetic}
We expand
\begin{equation}\label{eq:Phi_expand_general_section3_corrected}
  \Phi(x,y)=s^I(x)\Phi_I(y) \,,
  \qquad
  \partial_\mu\Phi=(\partial_\mu s^I)\Phi_I\,,
\end{equation}
where Greek indices $\mu,\nu$ label external spacetime coordinates, so $\partial_\mu$ denotes differentiation with respect to the external coordinates.
A generic deformation of $g_{mn}$ mixes a uniform rescaling of the internal space with deformations that change its shape at fixed volume.
To separate the overall volume modulus from the shape moduli of the $G_2$ metric, we therefore split off the overall volume by writing
\begin{equation}
  g_{mn}=\cV^{2/7}\tilde g_{mn},
  \qquad
  \Phi=\cV^{3/7}\tilde\Phi,
  \qquad
  \tilde\cV \equiv \int_X d^7y \sqrt{\tilde g_7} = 1 \,,
  \label{eq:volumeShapeSplit_section3_corrected}
\end{equation}
so that $\tilde\Phi$ depends only on the unit-volume deformations.

The part of the ten-dimensional Einstein--Hilbert term that contains $x$-derivatives of
the internal metric is
\begin{equation}
    (4\pi e)^{-1}\,\mathcal L_{s}^{R_7} = \frac{1}{8\mathcal{V}} \int_X d^7y \sqrt{g_7} \Bigl( g^{mp}g^{nq} - g^{mn}g^{pq} \Bigr) \partial_\mu g_{mp}\partial^\mu g_{nq}\,.
\end{equation}
Using \eqref{eq:volumeShapeSplit_section3_corrected}, this becomes 
\begin{equation}\label{eq:Ks_moduli_from_internal_metric_split_section3_corrected}
  (4\pi e)^{-1}\,\mathcal L_{s}^{R_7}
  =
  \frac37\,\partial_\mu\log\cV\,\partial^\mu\log\cV
  -\frac14\int_X \partial_\mu\tilde\Phi\wedge\tilde\star\partial^\mu\tilde\Phi \,.
\end{equation}
To rewrite this in terms of the unrestricted moduli $s^I$, note that
\begin{equation}\label{eq:dPhi_volumeShape_section3_corrected}
  \partial_\mu\Phi
  =
  \cV^{3/7}
  \left(
    \partial_\mu\tilde\Phi
    +\frac37\partial_\mu\log\cV\,\tilde\Phi
  \right) \,.
\end{equation}
Therefore 
\begin{align}\label{eq:L2decomp_section3_corrected}
  \frac{1}{\cV}\int_X \partial_\mu\Phi\wedge\star\partial^\mu\Phi
  &=
  \int_X
  \left(
    \partial_\mu\tilde\Phi
    +\frac37\partial_\mu\log\cV\,\tilde\Phi
  \right)
  \wedge
  \tilde\star
  \left(
    \partial^\mu\tilde\Phi
    +\frac37\partial^\mu\log\cV\,\tilde\Phi
  \right)
  \notag\\
  &=
  \int_X \partial_\mu\tilde\Phi\wedge\tilde\star\partial^\mu\tilde\Phi
  +\frac97\,\partial_\mu\log\cV\,\partial^\mu\log\cV \,,
\end{align}
while the cross term vanishes because $\tilde\cV=1$, hence
\begin{equation}\label{eq:crossTermVanishes_section3_corrected}
  \int_X \partial_\mu\tilde\Phi\wedge\tilde\star\tilde\Phi
  =
  3\partial_\mu\tilde\cV
  =0\,.
\end{equation}
Substituting \eqref{eq:L2decomp_section3_corrected} into
\eqref{eq:Ks_moduli_from_internal_metric_split_section3_corrected} gives
\begin{align}
  (4\pi e)^{-1}\,\mathcal L_{s}^{R_7}
  &=
  -\frac{1}{4\cV}\int_X \partial_\mu\Phi\wedge\star\partial^\mu\Phi
  +\frac34\partial_\mu\log\cV\,\partial^\mu\log\cV\cr
  &=
  -\frac14G_{IJ}\partial_\mu s^I\,\partial^\mu s^J
  +\frac34\partial_\mu\log\cV\,\partial^\mu\log\cV .
  \label{eq:Ks_moduli_naive_section3_corrected}
\end{align}

We now compute the second contribution, coming from the Weyl factor in the external metric.
Writing
\begin{equation}
  \widetilde g_{\mu\nu}=\cV^{-2}g_{\mu\nu} \,,
\end{equation}
the $d=3$ Ricci scalar transforms as
\begin{equation}\label{eq:WeylRicci_section3_corrected}
  R[\widetilde g]
  =
  \cV^2\Bigl(R_3+4\nabla_3^2\log\cV-2\,\partial_\mu\log\cV\,\partial^\mu\log\cV\Bigr) \,.
\end{equation}
Hence the Weyl contribution to the action is
\begin{equation}\label{eq:R3WeylContribution_section3_corrected}
\begin{split}
  \Delta S_{\mathrm{Weyl}}
  &=
  \frac{1}{2\kappa_{10}^2}\int d^{10}X\sqrt{-G_{10}}\,R[\widetilde g]\Big|_{\partial\cV} \\
  &=
  2\pi \int_{M_{2,1}}\!\sqrt{-g_3}\,
  \Bigl(4\nabla_3^2\log\cV-2\,\partial_\mu\log\cV\,\partial^\mu\log\cV\Bigr).
\end{split}
\end{equation}
The Laplacian term is a total derivative,
\begin{equation}
  \int_{M_{2,1}}\!\sqrt{-g_3}\,\nabla_3^2\log\cV
  =
  \int_{M_{2,1}}\!\partial_\mu\Bigl(\sqrt{-g_3}\,g_3^{\mu\nu}\partial_\nu\log\cV\Bigr),
  \label{eq:WeylTotalDerivative_section3_corrected}
\end{equation}
so, in the absence of boundary contributions,
\begin{equation}\label{eq:LWeyl_section3_corrected}
  (4\pi e)^{-1}\mathcal L_{\mathrm{Weyl}}
  =
  -\partial_\mu\log\cV\partial^\mu\log\cV \,.
\end{equation}
Adding \eqref{eq:Ks_moduli_naive_section3_corrected} and
\eqref{eq:LWeyl_section3_corrected}, the full Einstein-frame kinetic term is
\begin{equation}
  (4\pi e)^{-1}\,\mathcal L_{s}
  =
  -\mathcal G_{IJ}\partial_\mu s^I\,\partial^\mu s^J,
  \label{eq:G2moduliMetricPhysicalLagrangianapp}
\end{equation}
with
\begin{equation}
  \mathcal G_{IJ}
  =
  \frac14\,G_{IJ}
  +\frac14\,\partial_I\log\cV\,\partial_J\log\cV.
  \label{eq:G2moduliMetricPhysical}
\end{equation}
Next, using the standard variation of the $G_2$ volume,
\begin{equation}
  \partial_I\cV
  =
  \frac13\int_X \Phi_I\wedge\Psi \,,
  \label{eq:dVdSI_section3_corrected}
\end{equation}
this becomes
\begin{equation}\label{eq:G2moduliMetricPhysicalExplicitapp}
  \mathcal G_{IJ}
  =
  \frac14\,G_{IJ}
  +
  \frac{1}{36\,\cV^2}
  \left(\int_X \Phi_I\wedge\Psi\right)
  \left(\int_X \Phi_J\wedge\Psi\right) \,.
\end{equation}

\section{A contraction identity for the physical $G_2$ moduli metric}
\label{app:physical_metric_identity_corrected}

For the derivation of scalar potentials from a real superpotential, it is useful to complement the
completeness relation written purely in terms of $G_{IJ}$ by the corresponding identity for the
physical scalar metric of the $G_2$ moduli. We therefore start from the metric derived in
Section~\ref{subsec:3d_Einstein_frame},
\begin{equation}\label{eq:physicalG2metric_appendix}
  \mathcal G_{IJ}
  =
  \frac14 G_{IJ}
  +\frac14 v_Iv_J,
  \qquad
  v_I\equiv \partial_I\log\cV
  =
  \frac{1}{3\cV}\int_X \Phi_I\wedge\Psi
  =
  \frac13G_{IJ}s^J \,,
\end{equation}
where $\Psi=\star\Phi$ and $\Phi=s^I\Phi_I$.

The inverse metric can be obtained by a rank-one inversion. Since
\begin{equation}
  G_{IJ}s^Is^J
  =
  \frac{1}{\cV}\int_X \Phi\wedge\star\Phi
  = 7,
  \label{eq:sGs_equals_seven_appendix}
\end{equation}
one has
\begin{equation}
  v_I\,(G^{-1})^{IJ}v_J
  =
  \frac19\,G_{IJ}s^Is^J
  =
  \frac79.
\end{equation}
Applying the Sherman--Morrison formula to \eqref{eq:physicalG2metric_appendix} then gives
\begin{equation}\label{eq:physicalG2metric_inverse_appendix}
  \mathcal G^{IJ}
  =
  4G^{IJ}
  -\frac14\,s^Is^J.
\end{equation}

We now derive the corresponding contraction identity. For any 4-forms $\omega$ and $\eta$ lying in the truncated/harmonic subspace spanned by
$\{\Psi_I\}$, one finds
\begin{align}
  \mathcal G^{IJ}
  \Bigl(\int_X \Phi_I\wedge\omega\Bigr)
  \Bigl(\int_X \Phi_J\wedge\eta\Bigr)
  &=
  4\,G^{IJ}
  \Bigl(\int_X \Phi_I\wedge\omega\Bigr)
  \Bigl(\int_X \Phi_J\wedge\eta\Bigr)
  \notag\\
  &\hspace{1cm}
  -\frac14\,s^Is^J
  \Bigl(\int_X \Phi_I\wedge\omega\Bigr)
  \Bigl(\int_X \Phi_J\wedge\eta\Bigr)
  \notag\\
  &=
  4\,\cV\int_X \omega\wedge\star\eta
  -\frac14\,\Bigl(\int_X \Phi\wedge\omega\Bigr)
   \Bigl(\int_X \Phi\wedge\eta\Bigr),
  \label{eq:physicalGijCompleteness4form}
\end{align}
where in the last step we used the completeness relation based on $G^{IJ}$.
Equivalently, for any 3-forms $\alpha$ and $\beta$ in the truncated/harmonic subspace spanned by
$\{\Phi_I\}$,
\begin{equation}\label{eq:physicalGijCompleteness3form}
  \mathcal G^{IJ}
  \Bigl(\int_X \alpha\wedge\star\Phi_I\Bigr)
  \Bigl(\int_X \beta\wedge\star\Phi_J\Bigr)
  =
  4\,\cV\int_X \alpha\wedge\star\beta
  -\frac14\,\Bigl(\int_X \alpha\wedge\Psi\Bigr)
   \Bigl(\int_X \beta\wedge\Psi\Bigr)\,.
\end{equation}

Equations \eqref{eq:physicalGijCompleteness4form} and \eqref{eq:physicalGijCompleteness3form} are the identities needed when contracting derivatives of a real superpotential with the physical $G_2$ moduli-space metric. The second term subtracts the projection onto the overall volume direction and is absent if one uses only the auxiliary $L^2$ metric $G_{IJ}$.

\bibliographystyle{JHEP}
\bibliography{refs}

\end{document}